# Title: Measuring Very Low Radiation Doses in PTFE for Nuclear Forensic Enrichment Reconstruction

**Authors:** Rachel C. Connick[1], Charles A. Hirst[1], Kevin B. Woller[1], Julie V. Logan[1,2], R. Scott Kemp[1]*, Michael P. Short[1]*

**Affiliations:**

[1]Department of Nuclear Science and Engineering, Massachusetts Institute of Technology; Cambridge, MA 02139, USA.

[2]Air Force Research Laboratory Space Vehicles Directorate, Kirtland Air Force Base, Albuquerque, NM, USA.

*Corresponding author. Email: hereiam@mit.edu (M.P.S.), rsk@mit.edu (R.S.K.)

**Abstract:** Every country that has made nuclear weapons has used uranium enrichment. Despite the centrality of this technology to international security, there is still no reliable physical marker of past enrichment that can be used to perform forensic verification of historically produced weapons. We show that the extremely low radioactivity from uranium alpha emissions during enrichment leaves detectable and irreversible calorimetric signatures in the common enrichment gasket material PTFE, allowing for historical reconstruction of past enrichment activities at a sensitivity better than one weapon's quantity of highly enriched uranium. Fast scanning calorimetry also enables the measurement of recrystallization enthalpies of sequentially microtomed slices, confirming the magnitude and the type of radiation exposure while also providing a detection of tampering and a method for analyzing field samples useful for treaty verification. This work opens the door for common items to be turned into precise dosimeters to detect the past presence of radioactivity, nuclear materials, and related activities with high confidence.

**One-Sentence Summary:** Radiation signatures in common structural materials could verify whether a nuclear weapon has been made, and if so, how many.

## Main Text:

Accounting for weapons-usable fissile material is central to verifying international non-proliferation and arms-control agreements *(1–3)*. Enriched uranium is one of the two principal fissile materials for making nuclear weapons, and the only one for which forensic verification methods are not yet established. If produced under supervision, real-time accounting can take place by measuring material flows, a process known as safeguards. However, virtually all of the >1000 metric tonnes (MT) of weapon-usable uranium that exist today have been produced without safeguards. The International Panel on Fissile Materials estimates the uncertainty in past production at ±120 MT *(4)*, equating to ±5,000 nuclear weapons. For context, this uncertainty is over three times larger than the global number of deployed warheads allowed under the NewSTART disarmament treaty, making evident the need for better constraints on historical production *(5)*. Similarly, a number of countries, including Israel, North Korea, and South Africa, are believed to have produced nuclear weapons made with highly enriched uranium at undeclared facilities *(6–9)*, while Brazil, Iran, and South Korea have experimented with enriching uranium outside of safeguards *(10–12)*. Collectively, these cases illustrate the challenge international inspectors face when trying to ascertain whether all past production is known and accounted for—challenges that would be ameliorated if there were a way to forensically determine the amount of enriched uranium a facility has produced.

The original proposal of ”nuclear archaeology” for uranium enrichment formulated by Fetter, is based on the ratio of uranium-234 and uranium-235 in the waste streams of enrichment plants *(13)*. While this approach works in principle, it requires that all of the waste streams are available, none have been blended together, and that exceptionally precise measurements can be made *(14)*. These conditions cannot be met for many historical programs, and hiding waste containers from inspectors would be a trivially easy way to foil the effort. Two U.S. Department of Energy (DOE) laboratories revisited the challenge in 1993, but were unable to identify any readily reliable approaches *(3)*. The U.S. government effort was revived in 2013 following a growing need to find a solution to the problem. The Pacific Northwest National Laboratory proposed six different schemes, all of which depend on isotopic or elemental analysis of corrosion layers *(3, 15)* which require prohibitively delicate handling and are affected by a wide variety of environmental unknowns such as temperature and water vapor.

The signatures studied above are not tamper-proof and do not adequately retain historical enrichment signatures, amenable to simple inspection methods. The radiation damage resulting from the slow but steady alpha decay from the three isotopes of uranium ($^{234}$U, $^{235}$U, $^{238}$U) naturally present in uranium hexafluoride ($UF_6$) gas is a more permanent and robust signature. Each isotope has a well-documented half-life, alpha-particle energy, and, therefore, alpha-particle range; and the isotropic flux of alpha particle emissions will deposit energy in the surrounding material to a fixed, isotope-dependent maximum depth. As the isotopic composition of the $UF_6$ changes during enrichment, so too does the signature left behind from radiation damage, dominated by effects from the increasing abundance of $^{234}$U. This approach to nuclear forensics can be classified as a form of retrospective dosimetry, a discipline that has been applied to similar low dose-reconstruction applications, although no suitable methods exist yet for uranium enrichment verification.

Reconstruction of radiation doses in uranium-enrichment equipment is challenged by the overall low activity of $UF_6$. Even at 90% enrichment, doses are exceedingly low for most common structural materials. However, we demonstrate that the relatively light signature of alpha particle radiation damage at energies and doses expected in enrichment equipment after production of material for a single nuclear weapon can be reliably sensed and attributed to the radiation's type and energy by performing repeatable, calorimetric measurements of alpha-irradiated polytetrafluoroethylene (PTFE). Statistically differentiable changes in the enthalpy of recrystallization of irradiated PTFE relate changes observed *a posteriori* to radiation doses with sufficient resolution to distinguish weapons-grade and reactor-fuel-grade uranium production. The real-world applicability of this strategy is shown through a field-sampling method, enabled by fast scanning calorimetry (FSC) and micron thick sampled slices of PTFE gaskets, providing a more tamper-proof measurement of radiation signatures applicable to inspection of enrichment facilities for physical verification of compliance with non-proliferation treaties, or to reconstruct past enrichment activities.

## Radiation damage and its signatures

Charged particles, including alpha (α) particles, deposit energy in materials via two methods: nuclear and electronic stopping *(16)*. Nuclear stopping causes atomic displacements in a "damage cascade" *(17)*, a small (3–8 nm) ellipsoidal region of material which undergoes ballistic mixing, rapid heating, and cooling. This results in a permanent rearrangement of the atoms within the cascade *(18)*. Electronic stopping results in ionization and excitation *(16)*, which are largely reversible in metals but are permanent in covalently-bonded materials such as polymers. Very low-dose reconstruction for non-metals exists: electron paramagnetic resonance (EPR) has been used to turn teeth into extremely sensitive dosimeters *(19)* able to reconstruct radiation exposure down to single Gy levels. Additionally, Hayes and O'Mara proposed using thermal and optically stimulated luminescence of transparent crystalline minerals embedded in construction bricks to perform retrospective dosimetry *(20)*. However, these methods cannot be used in uranium-enrichment plants.

Uranium enrichment centrifuges are typically made of high-strength-to-weight ratio materials *(21)*, such as aluminum 7075-T6, maraging steels, or carbon fiber, and must be connected by piping compatible with the highly corrosive, fluorine-bearing $UF_6$ gas like aluminum *(22)* or passivated steel *(23)*. In addition, each centrifuge and associated piping may be connected by a sealing gasket. These can be made of compatible gasket materials like PTFE *(24)* or its chlorinated version, polychlorotrifluoroethylene (PCTFE) *(25)*. The signature of the historical presence of $UF_6$ on these materials must be reliably measured and understood to enable forensic enrichment reconstruction. The PTFE gasket material is a semi-crystalline fluoroplastic *(26)*, and suffers permanent chemical changes upon α-irradiation, mainly due to electronic excitation and subsequent generation of free radicals (unpaired electrons), ultimately resulting in irreversible changes to microstructure.

The effects of low doses of gamma, electron, and ion radiation on PTFE have been extensively studied *(27–30)*, with the threshold for severe degradation of material properties noted to occur at ~1000 kGy *(24)*, about two orders of magnitude greater than the doses in our work. Studies in prior literature typically use electron spin resonance (ESR) to detect stable free radicals, Fourier-transform infrared spectroscopy (FT-IR), ultraviolet visible absorption spectroscopy (UV-vis), or nuclear magnetic resonance (NMR) to detect bond densities; mechanical testing for degradation in material properties; and/or X-ray diffraction (XRD) or differential scanning calorimetry (DSC) to detect changes in crystallinity. The majority of studies on ion-irradiated PTFE focus on high doses, which are used to control surface wettability and adhesion *(31, 32)*. Very low dose ion-irradiation studies of PTFE are in the minority compared with low-dose studies of photons and electrons.

Forsythe and Hill review much of the early work on the mechanics of radiation interactions with PTFE *(26)*, citing the main effect as chain scission and a corresponding increase in crystallinity due to the reduction in molecular weight with radiation. Crosslinking can be achieved by irradiation above the melting temperature *(33, 34)*. More recent studies *(35–37)* have also shown that crosslinking can be achieved at room temperature under heavy-ion irradiation (including α-particles). However, this crosslinking is only appreciable at α-particle doses above1 MGy *(36)*, or in the narrow region of a Bragg peak *(37)* where the density of free radicals created by the energy deposited is higher. Below these doses, chain scission continues to be the dominant effect. In a similar material, poly(tetrafluoroethylene-co-hexafluoropropylene) (FEP), Yoshikawa et al. correlate the free radical distribution with depth to the radiation interactions of the Bragg peak (energy loss) using ESR at very low doses *(38)*, except at the end of the range where the higher density of production results in a higher degree of recombination of the free radicals. While this study contains doses comparable to our work, the high energy 24 MeV $He^{2+}$ ions and corresponding large damage volumes cannot be scaled down to meet our requirements, making their methods inapplicable to ours.

Despite the wealth of literature available, there is still a dearth of data for very low dose (<10 kGy) effects in PTFE from ions like α-particles, which is necessary to define the limits of sensitivity in verifying uranium enrichment. For this data we turn to DSC, which can measure bulk thermal properties and avoid surface effects. Furthermore, DSC is fairly robust in the geometries it can handle, especially since the advent of micro- and nanocalorimetry and fast scanning calorimetry (FSC) devices, as long as good thermal contact can be achieved. Because a 4.5 MeV α particle only penetrates ~20μm into PTFE, FSC permits direct characterization of the radiation-depth profile as a second physical signature of enrichment by microtome sectioning of the irradiated layer, which provides information related to the isotopic source of the radiation. Further sampling into the gasket beyond the 20 μm ion range allows for a built-in control in the form of unirradiated material which underwent identical processing and history due to its tens-of-microns coincident location. PTFE has been considered as a candidate for dosimetry applications before *(39–41)*, albeit in gamma radiation applications, which is characterized by larger, more uniform damage volumes that are not directly comparable to ion irradiation. We must show that PTFE can be used in retrospective dosimetry as well, without having the ability to control factors such as geometry or manufacturing parameters beforehand, as can be done with traditional dosimetry.

Now we must answer the question of whether such signatures can be used as dosimeters to detect the dose from α-exposure and thus recreate the operating history of an enrichment plant. Isotopes of uranium spontaneously decay with known half-lives and decay energies, and are summarized in Table 1. Here we also compare the relative contribution of each isotope to the total activity calculated for low enriched (4%) and weapons grade (93%) $^{235}U$ levels. The minor isotope $^{234}U$ contributes the most activity at all enrichment levels, particularly so in highly enriched $UF_6$, and dominates the dosimetric signature. Should a specific measurement be conclusively linked to an energy deposition in kGy, then the inverse problem of reconstructing radiation dose, and thereby α-fluence and $UF_6$ throughput, would be within reach, assuming knowledge of the enrichment plant geometry.

## Modeling isotopic mix during $^{235}U$ enrichment

We start by estimating the α-particle flux incident on the inner surface of centrifuge-based enrichment pipework to distinguish between three cases: 5% low enriched uranium (LEU) typical of nuclear fuel, 20% enriched uranium as the defined limit to highly enriched uranium (HEU) *(1)*, and 90% weapons-grade enriched uranium which has few applications aside from nuclear weapons (42). The expected flux in a $UF_6$-containing pipe, assuming all emitted α-particles hit the surrounding PTFE with negligible self-attenuation in the $UF_6$ gas, can be calculated using Equation 1,

$$\Phi = \frac{\ln(2)}{4} D_{pipe} N_{tot} \sum_i \frac{f_i}{t_{1/2,i}} \tag{1}$$

Where $D_{pipe}$ is the assumed pipe diameter, $N_{tot}$ is the molecular density of the $UF_6$ gas in the pipe, $f_i$ is the fraction of each isotope in the mix, and $t_{1/2,i}$ is the half life of each isotope.

The α fluence is dominated not by $^{235}U$, but by the minor isotope $^{234}U$. The abundance of this isotope in uranium varies slightly across ore bodies, and will further vary depending on the characteristics of centrifuges in use. To estimate $^{234}U$ abundance at various $^{235}U$ enrichment levels, we construct a semi-empirical model based on basic centrifuge theory and informed by published experimental data *(43–46)*. A value of 1% $^{234}U$ at an enrichment at 93% $^{235}U$ matches both experimentally measured *(43–45)* and simulated *(46)* data. More details of the development of this model can be found in the supplementary materials. Combining the semi-empirical model of $^{234}U$ enrichment with Equation 1, yields the expected fluxes and doses shown in Figure 1.

The decay of $^{234}U$ is dominated by two α lines, 71% are at 4.774 MeV and 28% at 4.722 MeV. SRIM simulation *(47)* yields an α-particle range of 21.3 μm in PTFE (density 2.2 g/cm$^3$) for 4.7 MeV α particles, using the built-in compound-correction factors *(48)*. This corresponds to an energy deposition in the irradiated volume of 1.6 kGy, or J/g. While this level of energy deposition is quite low, its effects should be measurable in DSC, based on prior literature *(30)*.

**Experimental parameters maximize signal-to-noise ratio**

We show that the material PTFE is sufficiently sensitive to the radiation doses of interest, and that DSC can resolve the effects of these changes. Figure 2 shows the PTFE sampling methods used for the DSC and FSC experiments, where thin, flat samples were key. We test a large matrix of samples irradiated to the doses of interest, identified in Figure 2A. The relative ease of performing ion-beam irradiations using a 1.7 MV tandem accelerator facilitated relatively high sample numbers to improve the statistics of our results. By sampling from a thin film and maximizing the sample diameter, the signal from the irradiated volume is maximized.

While conventional DSC measurements capture the cumulative effect of the radiation, we also show the sensitivity of PTFE when coupled with the higher heating rates and smaller sample size enabled by FSC and nanocalorimetry *(49, 50)* to measure the depth profile of the radiation effects. We did this at $1 \times 10^{11}$ α/cm$^2$ with five individual samples. By microtome-sectioning nominally 1 μm disks of relatively large (250 μm) diameter, each slice can be measured for a comparable result to the conventional DSC tests.

The depth profile of the monoenergetic, collimated $He^{2+}$-ion beam will differ from the isotropic α-particle source resulting from $UF_6$ gas and the latter will require greater signal-to-noise to resolve between different α particles (isotope decays). FSC also enables the removal of tiny amounts of material from an enrichment facility, so small that they could be considered non-destructive, allowing treaty verification inspectors to obtain hard-to-spoof primary information directly from an enrichment plant. Details of the materials and methods of all experiments can be found in the supplemental material.

**Enthalpy measurements are statistically distinguishable**

Figure 3 shows the DSC results, with recrystallization curves from each sample to show repeatability, as well as the extracted data with error bars for the recrystallization enthalpy $\Delta H_{cryst}$. After subtracting a spline-type baseline, described further in the supplemental material, the DSC data shows reliable behavior between fluence and recrystallization. We quantify this by integrating the baseline-corrected DSC data to determine $\Delta H_{cryst}$. The bands around the average values are one and two standard deviations for each set of irradiated samples. The magnitude of the variance in $\Delta H_{cryst}$ is invariant with fluence indicating that this uncertainty results from inherent variation between the unirradiated samples (see supplemental material). Some expected $\Delta H_{cryst}$ values are extrapolated for 5% (reactor fuel-grade), 20% (borderline HEU), and 90% (weapons-grade) enriched uranium to demonstrate how $\Delta H_{cryst}$ can be related back to enrichment in Figure 3B.

The observed trends in $\Delta H_{cryst}$ are in agreement with those seen throughout the irradiated PTFE literature *(30, 36, 37, 51)*. Qualitatively, as the dose increases, the enthalpies also increase in magnitude, indicating a higher degree of crystallinity *(51)*. This increase in crystallinity is attributable to chain scission reducing the average molecular weight in the material *(29)* and enhancing crystallinity. Although ion (α-particle) radiation has also been observed to result in branching or crosslinking *(36, 37)*, the dose range in this study lies far below the threshold for such effects to dominate (about 1 MGy *(36)* compared to 1–10 kGy here). This dose window between 10 kGy and 1 MGy corresponds to 100 years of weapons-grade enrichment in gas centrifuges, but potentially opens the door to quantification of higher-pressure enrichment processes such as historical gaseous-diffusion plants used to produce most of the weapon-grade uranium that exists today. Direct, absolute comparison with the cited literature, however, is difficult due to effects of the initial condition of the PTFE *(29, 52–54)*, type of radiation *(37)*, and possibly irradiation atmosphere *(27, 52, 54, 55)* or dose rate *(56)*. Furthermore, at least part of the difference can be attributed to differences in data processing, as the definition of the base-line parameters will affect the absolute enthalpies calculated. Nevertheless, we are consistent with the reported orders of magnitude for enthalpies and their changes with radiation for similar doses *(30, 36, 37, 51)*, and our control surface methodology (see supplementary information) minimizes the effects of the highly localized differences in PTFE manufacture resulting from the sheet skiving process.

Figure 3B confirms the high sensitivity of PTFE to radiation, and that this sensitivity meets the requirements for detection of illicit activity set by Figure 1. Based on the scenario described by the enrichment model and assuming a quadratic trend in the irradiated sample data, 90% enriched uranium (weapons-grade) correlates to a $\Delta H_{cryst}$ of 27.3 J/g in this PTFE. Comparatively, production of 5% (typical fuel for a nuclear power) or even 20% enriched uranium (legal limit for low enriched uranium production) correlate to much lower enthalpies (20.5 J/g and 22.9 J/g, respectively), which themselves are distinguishable from the unirradiated PTFE (18.9 J/g). The p-values comparing each of these results to the others are less than <0.001, indicating that each of these measurements should be significantly different from the others, even with the relatively small sample sizes of 10–20 measurements per fluence.

The foregoing results are derived strictly from the total dose received by each sample. If only one sample is obtained, there is a degeneracy: the result could be attributable either to a small amount of high-enriched uranium or to a large amount of low-enriched uranium. In most cases other characteristics of the enrichment process can be used to differentiate between these two scenarios. For example, about 280 times more 5% enriched uranium would have to have been produced to give the same fluence as 90% enriched uranium, which if true would imply a large amount of unaccounted-for LEU. However, with a second sample at another point in the centrifuge cascade, the degeneracy can be broken. A total of $N$ samples can be used to reconstruct up to $N-1$ enrichment campaigns, each having a different enrichment end point.

## Depth-profile reconstruction using FSC

Figure 4 shows FSC measurements on five sets of sequentially microtome-sectioned slices that map the radiation effects of the highest irradiation from the DSC study ($1\times10^{11}$ α/cm$^2$) into the depth of the sample. The uncertainties shown are 1 and 2 standard deviations for a 2 μm moving average, since the number of measurements at each coordinate varies. Unlike with the conventional DSC measurements, the variance in this data is most likely attributable to different measurement conditions, as achieving consistent thermal contact between micron-size samples and the FSC sensor is challenging. The difficulty in microtome-sectioning up to 32 uniform, consecutive slices also contributes to the variation observed in the data, especially as accumulated error in the depth coordinate muddles the transition between the irradiated and unirradiated regions. These measurements also have an additional source of uncertainty in the need to calculate a sample mass using the measured heat capacity *(57)* rather than making an independent measurement of mass, although this is more likely to introduce a systematic, rather than random, error.

The magnitudes of $\Delta H_{cryst}$ measured by FSC differ from those measured in the conventional DSC. This is due to the much faster cooling rate utilized in FSC compared to DSC (600 °C/s versus 0.17 °C/s, respectively). Bosq et al. *(58)* demonstrated decreasing enthalpies of melting $\Delta H_{melt}$ after cooling at varying rates in PTFE, showing the $\Delta H_{melt}$ after cooling at 500°C/s to be about 75% of that after cooling at 0.1°C/s, due to quenching of the samples. Our DSC experiments studied samples from a 50 μm-thick film, more than twice the range of 4.5-MeVα-particles in PTFE. Taking this into account, the measured value at the $10^{11}$ α/cm$^2$ fluence in the conventional DSC result (Figure 3B) can be thought of as the average of the measurements of a 20 μm irradiated section and a 30 μm unirradiated section. As such, assuming the unirradiated piece would have the same $\Delta H_{cryst}$ as that measured for the unirradiated control samples, the measurements for each piece might be expected to come out to ∼45 and ∼18 J/g respectively, which can be compared to the FSC measurements. The FSC measurements in the irradiated and unirradiated regions are closer to ∼30 and ∼12 J/g, about 66% of the expected values, attributable to quenching by the faster cooling rate. Other small differences may come from differences in data-processing, such as the choice of baseline function when integrating to get $\Delta H_{cryst}$.

Based on the SRIM-calculated Bragg curve (stopping power), which predicts the range of 20μm, the FSC data can be separated into two regions. Visually, the predicted range matches the behavior of the data well. This behavior is in accordance with Yoshikawa et al. *(38)* and Gowa et al. *(59)*, who have been able to map radiation effects to a heavy-ion Bragg curve with high energy particles in stacked films using ESR and FT-IR in similar materials (fluorinated ethylene propylene (FEP) and ethylene tetrafluoroethylene (ETFE), respectively). Pugmire et al. *(36)* also measured the FT-IR depth profile in stacked films of PTFE with a more comparable energy to ours, at a dose of 100 kGy and a resolution of 5 μm. One element our data does not show when compared to the Bragg curve is a significant peak in $\Delta H_{cryst}$ at the end of the irradiated region, instead exhibiting more of a shelf in the irradiated region. As previous authors have shown, ion radiation can deposit sufficient energy in a small volume and time at the end of its range that either crosslinking can be achieved *(36)* or the high density of free radicals produced have a greater chance of recombining with each other *(38)*. Either of these effects would reduce the impact of the deposited energy on the measured $\Delta H_{cryst}$. While the transition between the regions is also not as sharp as might be expected, it is possible that this more gradual transition is attributable to how the uncertainty in the actual depth of each slice effectively compounds with every sequential slice during the microtome-sectioning process.

Thus, we show how a measurement of the depth of the radiation can be extracted from this data, through comparison to a simulated profile. By taking this depth-profile measurement of a sectioned sample, a second characteristic of the radiation is measured. While it is possible to recreate the cumulative dose effect in a sample with alternative sources of radiation, recreating the actual profile is not trivial. A monoenergetic, uni-directional helium-ion beam will leave a much different profile than an isotropically emitting gas source like a decaying $UF_6$ gas (see supplementary information). Therefore, this profile-measurement technique illustrates a process by which spoofing could be detecting.

This work shows that low-dose radiation signatures are measurable, and attributable to their originating radioisotope, in a fairly ubiquitous material—opening the door for common items to be turned into precise dosimeters to detect the past presence of radioactivity, nuclear materials, and related activities. While PTFE is an ideal case, with a pre-existing body of literature and well-characterized behavior, the same methodology can be applied to many more materials. This may be the start of a new domain of nuclear forensic sciences with immediate application to nuclear arms-control efforts.

**Acknowledgments:**

This work is supported in part by the Consortium for Verification Technology (CVT) under Department of Energy National Nuclear Security Administration Award DE-NA0002534, and the consortium for Enabling Technologies and Innovation (ETI) under Department of Energy National Nuclear Security Administration Award DE-NA0003921. This material is based upon work supported by the National Science Foundation under Grant No. DMR-1654548. Any opinions, findings, and conclusions or recommendations expressed in this material are those of the author(s) and do not necessarily reflect the views of the National Science Foundation.

Disclaimer: This report was prepared as an account of work sponsored by an agency of the United States Government. Neither the United States Government nor any agency thereof, nor any of their employees, makes any warranty, express or implied, or assumes any legal liability or responsibility for the accuracy, completeness, or usefulness of any information, apparatus, product, or process disclosed, or represents that its use would not infringe privately owned rights. Reference herein to any specific commercial product, process, or service by trade name, trademark, manufacturer, or otherwise does not necessarily constitute or imply its endorsement, recommendation, or favoring by the United States Government or any agency thereof. The views and opinions of authors expressed herein do not necessarily state or reflect those of the United States Government or any agency thereof.

**Author contributions:**

Conceptualization: RSK, MPS, RCC, CAH

Methodology: RCC, MPS, RSK

Investigation: RCC, JVL, KBW

Visualization: RCC, MPS

Funding acquisition: MPS, RSK

Project administration: MPS, RSK

Supervision: MPS, RSK

Writing – original draft: RCC, MPS, RSK

Writing – review & editing: RCC, CAH, KBW, JVL, RSK, MPS

**Competing interests:** Authors declare that they have no competing interests.

**Data and materials availability:** All data, code, and materials used in the analysis are available through a GitHub repository, DOI:10.5281/zenodo.4872301.

## Supplementary Materials

Materials and Methods

Supplementary Text

- $^{234}U$ Enrichment Model
- SRIM simulations and damage profiles
- DSC settings and parameters
- Location effects in PTFE film stock
- FSC settings and parameters
- Data processing for DSC and FSC
- Outlier identification for FSC
- Mass calculation for FSC
- Microtome sectioning technique

Figs. S1 to S12

Table S1

References (60–68)

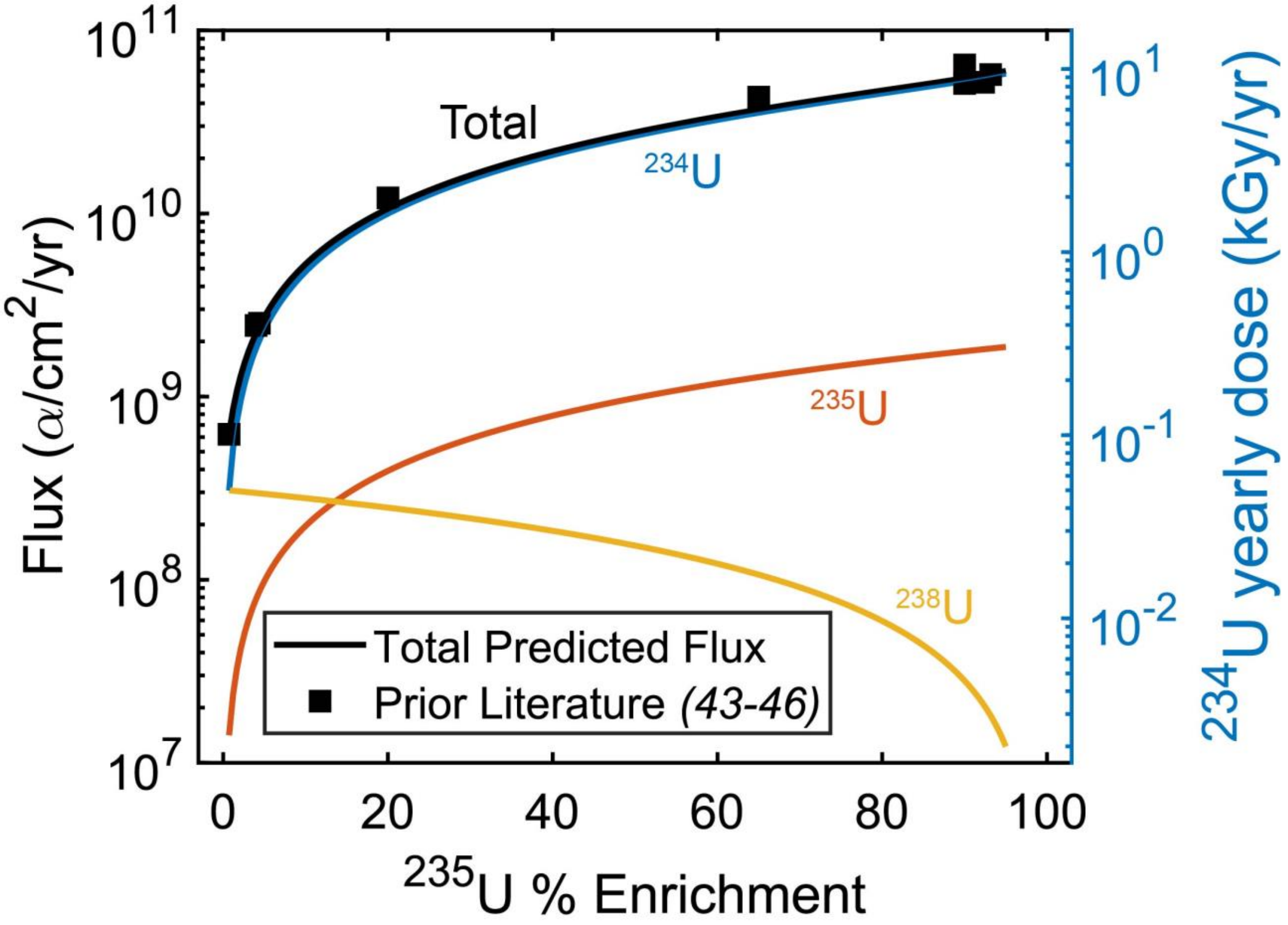

**Fig. 1. Flux and yearly dose vs $^{235}$U enrichment based on a semi-empirical model of $^{234}$U enrichment and Equation 1.** As indicated by the overlap of the total flux curve and its $^{234}$U component, $^{234}$U comprises almost the entirety of the α particle flux due to its relatively short half-life, even though it is estimated to make up about 1% of the uranium at very high enrichments. For this reason, the dose estimate is simplified by calculating only the $^{234}$U contribution.

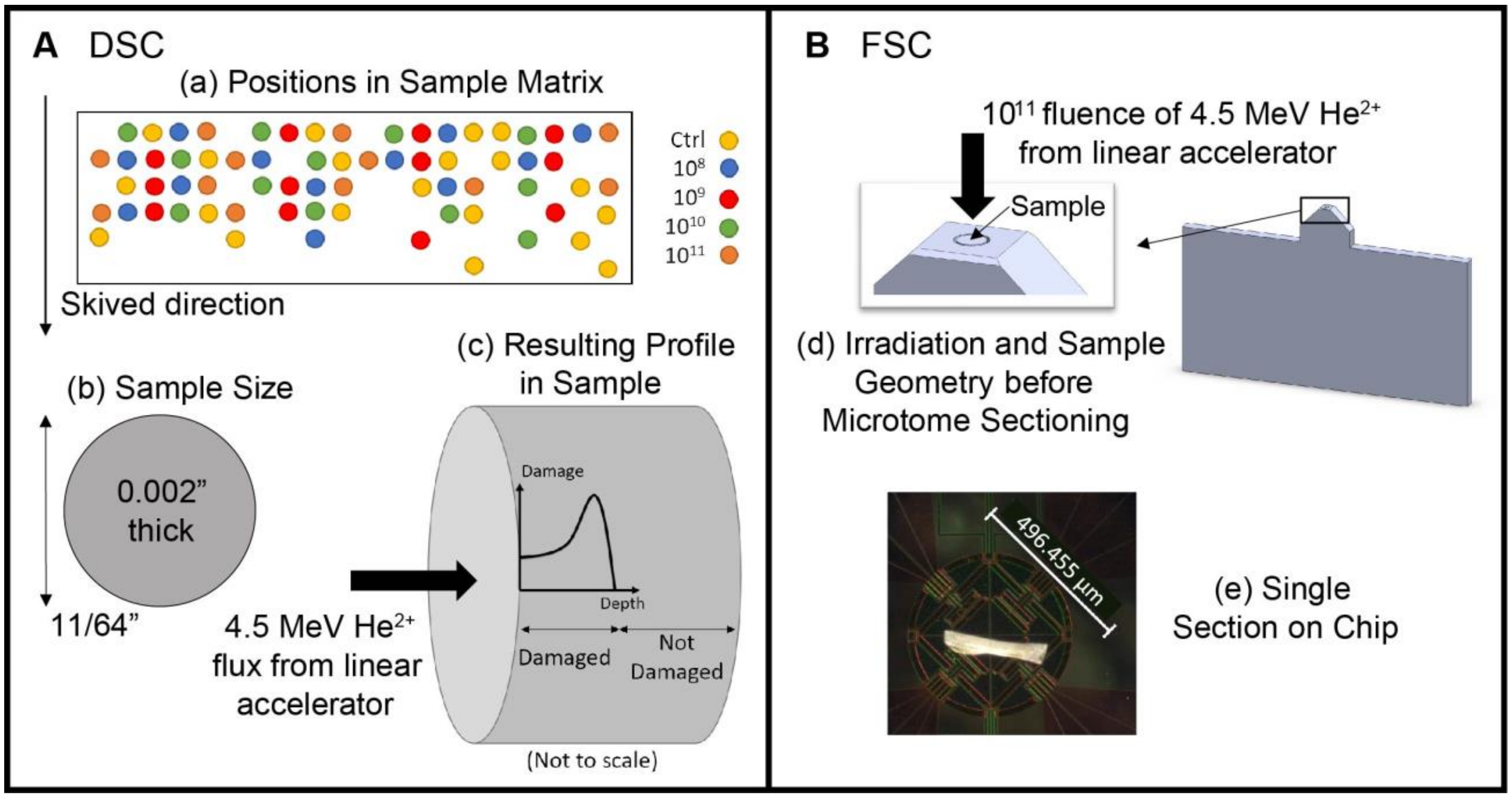

**Fig. 2. Sampling method and irradiation setup for (A) DSC and (B) FSC.** (a) Locations of each DSC sample were recorded to check location dependencies in the stock material through the unirradiated control sample measurements. (b) Sample dimensions with maximized radius and minimized thickness. (c) The damage profile from the accelerator beam does not fully penetrate the sample. (d) Irradiation setup by accelerator beam for microtome-sectioning and FSC measurements. (e) One single section on FSC chip sensor after measurement. Some samples curled up during the measurement.

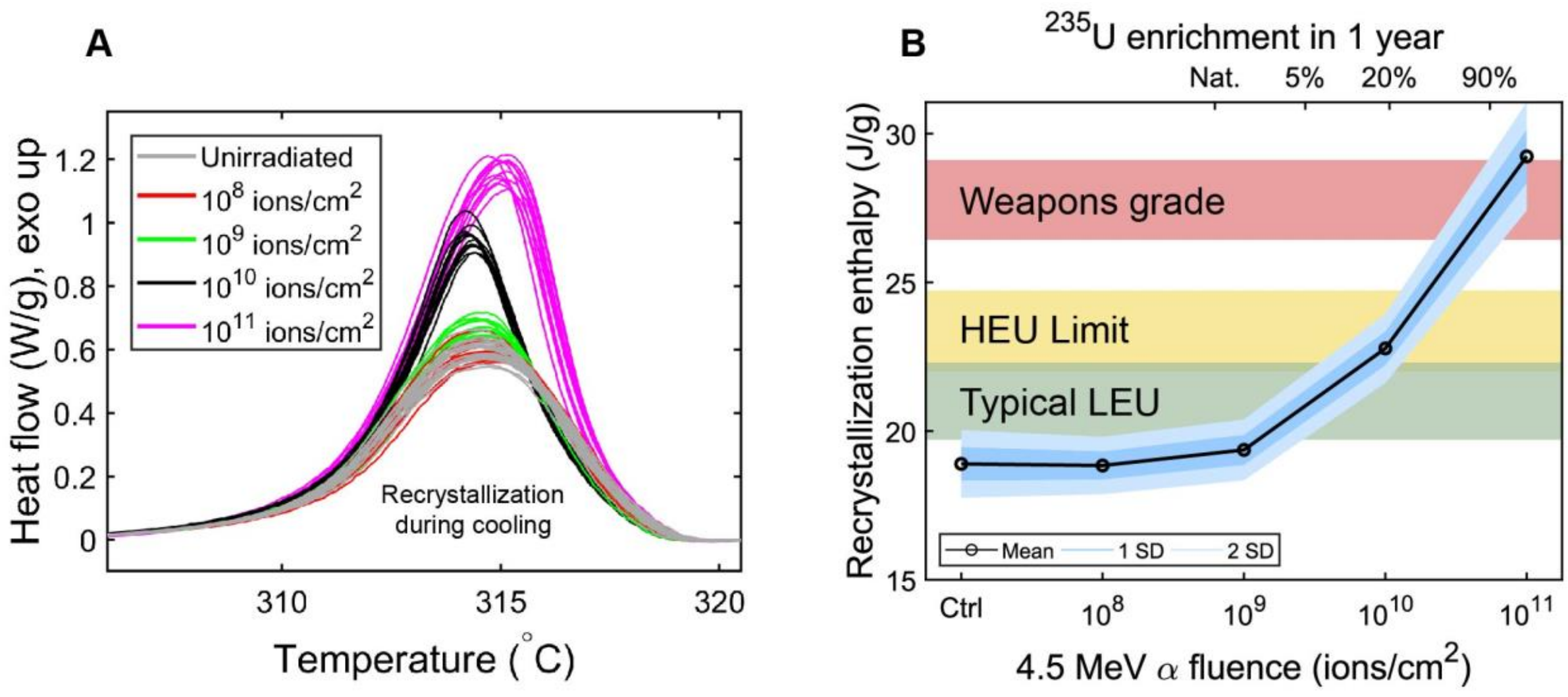

**Fig. 3. DSC measurements with varying fluence. (A)** Post-processed data clusters well with fluence. The recrystallization peaks at each fluence are characterized by their enthalpies, as reported in (B), with 1 and 2 standard deviation uncertainties. The lower horizontal axis in (B) has a non-logarithmic break to display the control sample measurements, which were exposed to no fluence. Fluences correspond to expected 1-year exposure from varying enrichments, which has been projected on the upper horizontal axis of (B), based on the model in Figure 1. Assuming a quadratic trend in the irradiated samples, the estimated 95% confidence intervals of enthalpies for 1-year exposure of 5% (typical reactor fuel LEU), 20% (legal limit for HEU), and 90% (approximately weapons-grade uranium) are displayed.

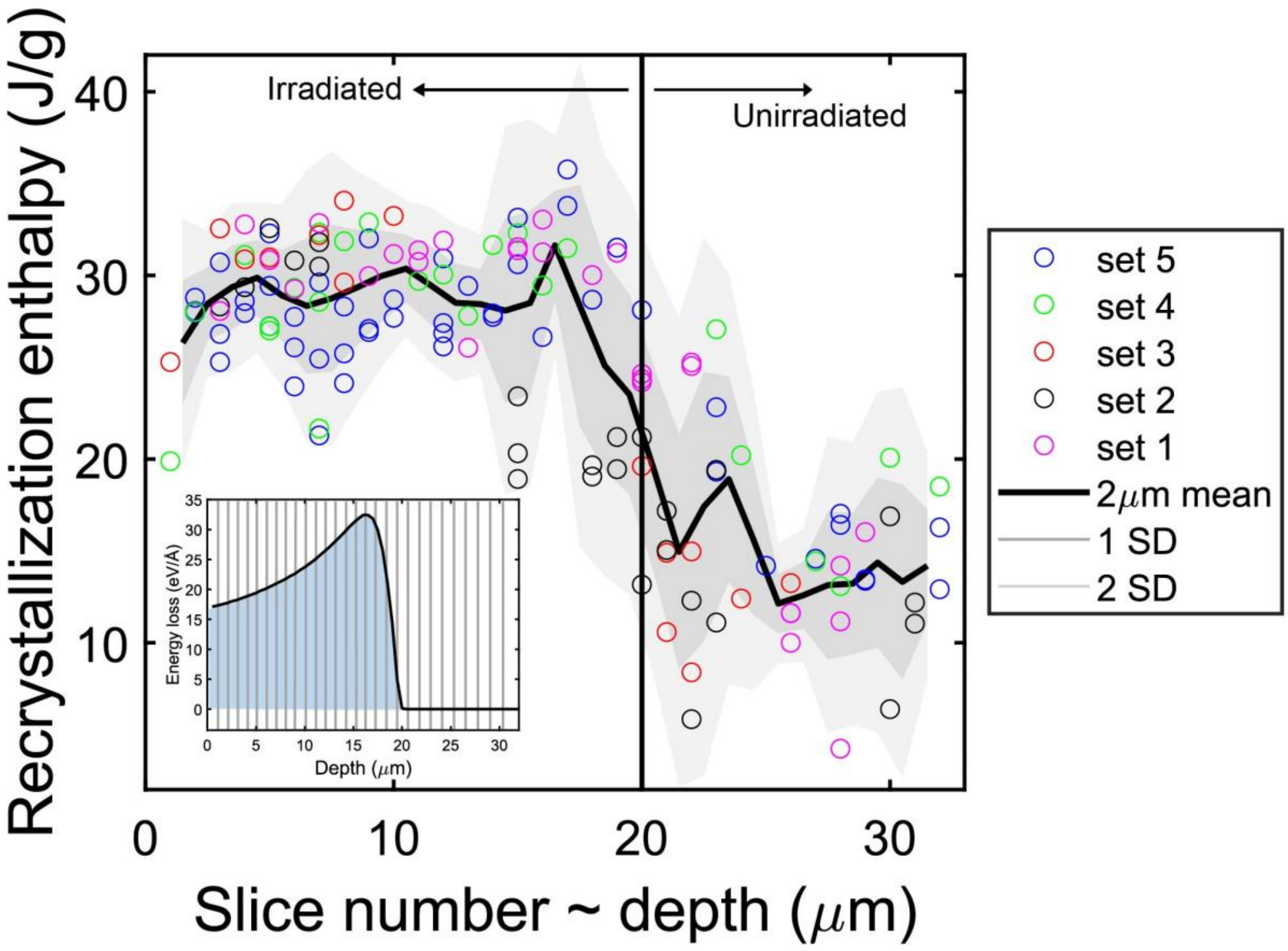

**Fig. 4. Depth-dependent irradiation effects on recrystallization measured by FSC.** The data is reported with a 2-μmmoving average with the associated 1 and 2 standard deviations. The inset shows the Bragg curve (stopping power) predicted by SRIM, where the area under the curve is what is measured by DSC, and FSC is measuring along the curve. Qualitatively, the 20 μm SRIM-predicted range separates the data into irradiated and unirradiated regions well. While the Bragg curve shows a sharper transition compared to the FSC data, this discrepancy can be explained by the compounding uncertainty on the depth-slice number relationship due to the sequential nature of the microtome-sectioning.

| **Isotope** | **$t_{1/2}$ (yr)** | **$E_\alpha$ (MeV)** | **R (μm)** | **$\%\phi_{LEU}$** | **$\%\phi_{HEU}$** |
|---|---|---|---|---|---|
| $^{234}U$ | $2.5 \times 10^5$ | 4.7 | 21.3 | 84.72% | 96.8% |
| $^{235}U$ | $7.0 \times 10^8$ | 4.4 | 18.8 | 3.44% | 3.17% |
| $^{238}U$ | $4.5 \times 10^9$ | 4.2 | 17.7 | 11.78% | 0.03% |

**Table 1. Isotopes of uranium naturally found in $UF_6$, their half lives, their alpha decay energies, and their flux contributions at chosen enrichments.** $E_\alpha$ - Most likely alpha particle energy; R – Range in PTFE; %Φ - percent contribution to flux at 4% (LEU) and 93% (HEU) enrichments.

# Supplementary Materials for

## Title: Measuring Very Low Radiation Doses in PTFE for Nuclear Forensic Enrichment Reconstruction

**Authors:** Rachel C. Connick[1], Charles A. Hirst[1], Kevin B. Woller[1], Julie V. Logan[1,2], R. Scott Kemp[1], Michael P. Short[1]*

**Affiliations:**

[1]Department of Nuclear Science and Engineering, Massachusetts Institute of Technology; Cambridge, MA 02139, USA.

[2]Air Force Research Laboratory Space Vehicles Directorate, Kirtland Air Force Base, Albuquerque, NM, USA.

Correspondence to: hereiam@mit.edu

**This PDF file includes:**

## Materials and Methods

Materials and Specimen Fabrication

Polytetrafluoroethylene (PTFE, Teflon) was purchased in two forms from McMaster Carr: 0.002” (50μm) skived film stock for regular DSC measurements (part no. 8569K12) and 1/32” (0.8 mm) sheet stock for FSC measurements (part no. 1063T11). For irradiation and regular DSC experiments, the film was punched into 11/64” (4.4 mm) diameter discs using a hammer-driven hole punching tool. The masses of these discs were on average 1.72 mg with a standard deviation of 0.02 mg, indicating fairly high repeatability using this method. For the FSC experiments, small cross-sections of the sheet stock were irradiated, and then sequentially sectioned by a Leica EM UC7 Ultramicrotome with cryostage attachment in increments of 1 μm. The microtome environment was kept between -25 and 0 degrees Celsius.

Helium Ion Irradiation

Irradiations were carried out in the Cambridge Laboratory for Accelerator-based Surface Science (CLASS) 1.7 MV tandem ion accelerator at MIT. For the regular DSC experiments, the film discs were irradiated with 4.5 MeV monoenergetic and unidirectional $He^{+2}$ ions to four fluences: $1\times10^{8}$, $1\times10^{9}$, $1\times10^{10}$, and$1\times10^{11}$ ions/cm$^2$. Samples were secured against the sample holder with a tungsten mesh possessing 92% transmission area, leaving 8% of the sample faces unirradiated. The beam was broadened to fill an area of 15 mm by 15 mm at the target location with uniform flux density as measured by a beam profile monitor downstream from the sample holder. With this method, it was possible to irradiate up to 9 samples in a single batch to the specified ion fluence. The ion beam current measured on the target holder during the irradiations was integrated for collected charge. The intended ion fluences were achieved by only allowing the exposure to carry on until the intended charge was collected. Ion impact leads to secondary electron emission, which generate uncertainty in the collected charge. However, the sample holder was surrounded by a faraday cage biased to -500V with an opening for the beam to pass through. Thus, the intended ion dose was accurate to within ±15%. The resulting matrix for the DSC experiments contained 12 samples at each fluence, as well as 20 unirradiated control samples. For the FSC experiments, the 5 pieces of the PTFE sheet were irradiated together in cross section with 4.5 MeV $He^{+2}$ ions to $1\times10^{11}$ ions/cm$^2$. All irradiations were performed at room temperature in a vacuum of $1\times10^{-6}$ Torr or less.

Differential Scanning Calorimetry (DSC)

The recrystallization enthalpies of the film samples were measured using a TA Instruments Discovery DSC. Experiments were performed in a 50 mL/min flowing 99.999% pure nitrogen atmosphere with Tzero aluminum pan and lid sample holders. Both heating and cooling were conducted at rates of 10°C/min between 250 and 375°C. 5-minute isotherms between each segment were used to allow enough time for thermal equilibrium to be reached between measurements. The cycle was run twice, to compare only the repeatable and reversible changes between samples, as captured in the second cycle. Samples were each weighed before DSC measurement, and the mass of each sample was used to normalize DSC power output in J/g.

Fast Scanning Calorimetry (FSC)

For the FSC experiments, the irradiated samples were microtome-sectioned, after which the recrystallization enthalpies along and beyond the depth of the radiation (as set by the range of the alpha particles in PTFE) were measured in a Mettler Toledo Flash DSC 1 with UFS1 chip sensors. The atmosphere was 99.999% pure argon flowing at 50 mL/min. The samples were measured from 250 to 375°C at heating and cooling rates of 500 K/s with 0.1 s isotherms in between. Four cycles were measured for each sample. Masses for each sample were calculated from the measured total heat capacity away from the melting and recrystallization peaks, and were compared to temperature-dependent specific heat capacity data from literature *(60)*.

**Supplementary Text**

$^{234}$U Enrichment Model

The short half-life (246,000 years) of $^{234}$U may suggest that it cannot be present in natural uranium, however it is a natural daughter product of $^{238}$U decay produced via the following nuclear reactions

$$^{238}U \underset{t_{1/2}=4.468\times10^{9}\,y}{\overset{\alpha^{+2}}{\rightarrow}} {}^{234}Th \underset{t_{1/2}=24.1\text{ d}}{\overset{\beta^{-}}{\rightarrow}} {}^{234}Pa \underset{t_{1/2}=6.7\text{ h}}{\overset{\beta^{-}}{\rightarrow}} {}^{234}U \qquad (S.1)$$

Because of the relatively high activity of $^{234}$U in enriched uranium due to its relatively short half-life, it is important to be able to estimate the enrichment of $^{234}$U as a function of $^{235}$U enrichment, as it constitutes the vast majority of α flux in an enrichment centrifuge. While it is possible to directly simulate the enrichment of $^{234}$U *(46,61)*, we chose to use basic enrichment theory *(61–64)* to develop a simple order-of-magnitude estimator for $^{234}$U enrichment as a function of $^{235}$U enrichment. As we show in this section, the result varies greatly according to the simplifying assumptions, and we ultimately used a semi-empirical method to align our model to available data *(43–45)* and simulations *(46)*.

The goal of the following calculations is to estimate the enrichment of $^{234}$U as a function of the enrichment of $^{235}$U. Because $^{234}$U is lighter than $^{235}$U, it will enrich relatively more than $^{235}$U, as confirmed by the separation factors determined below. The general approach to predict this $^{234}$U enrichment is as follows: given a desired $^{235}$U enrichment, the number of stages necessary to achieve that enrichment can be calculated. From there, the $^{234}$U enrichment is calculated, assuming the same number of enrichment stages. If $f_i$ is the atomic (and molecular) fraction of each isotope, in a three-component system, the relationship $1 = f_{238} + f_{235} + f_{234}$ should always hold true.

Each stage of centrifuges in the cascade is assumed to have the same separation factor $\alpha_i$ for each isotope $i$. $\alpha_i$ is defined as the ratio of the relative abundance of $i$ in the enriched stream to that in the depleted stream from each centrifuge *(62,64)* as follows:

$$\alpha_i = \frac{\left({f_i}/{f_{238}}\right)_{enriched}}{\left({f_i}/{f_{238}}\right)_{depleted}} \qquad (S.2)$$

By convention, the abundances are considered relative to the ”key” isotope, usually $^{238}$U *(62,64)*. The typical value for $\alpha_{235}$ ranges from 1.2–1.6 *(65)*, where we will use 1.4 for all the following calculations. The relationship between separation factors for different isotopes can be calculated from the relationships developed by Von Halle *(63)* or Wood *(64)*:

$$\alpha_i\,(Von\ Halle) = (\alpha_{235})^{(238-i)/(238-235)} \qquad (S.3)$$

$$\alpha_i\,(Wood) = \frac{238-i}{238-235}(\alpha_{235}-1) \qquad (S.4)$$

From these equations, $\alpha_{234}$ is 1.566 (Von Halle) or 1.533 (Wood). Wood's approximation will be used in our calculations, because it is derived specifically for centrifuge-type enrichment. Thus, using these separation factors, for a cascade with $N_s$ stages to enrich from a given input to the desired output of enriched $^{235}$U, the overall separation factor equation for an entire cascade becomes

$$\alpha_i^{N_s} = \frac{\left({f_i}/{f_{238}}\right)_{output}}{\left({f_i}/{f_{238}}\right)_{input}} \tag{S.5}$$

We will assume the input in these calculations is natural uranium with 0.72 at. % $^{235}$U, 0.0055 at. % $^{234}$U, and the remainder as $^{238}$U. With Equation S.5 we can use a range of $^{235}$U enrichments to calculate the number of stages necessary to achieve those enrichments, and then use Equation S.5 with Equation S.4 to get the enrichment of $^{234}$U after that same number of stages. The mathematical issue with this approach is how to deal with the fraction of $^{238}$U, which comprises the balance. If we use the full constraint $1 = f_{238} + f_{235} + f_{234}$, then the $f_{234}$ variable becomes impossible to algebraically isolate in a simple manner, and the model may not have any real solutions above about $f_{235}$ = 90%, which is described in further detail later on. In the following calculations, we demonstrate the different outcomes based on a variety of assumptions in detail. Additionally, we present a semi-empirical model that can be used to adjust the model to match the expected values. Each of these models are presented in Figure S.1 for comparison.

Assuming $f_{234}$ and $f_{235}$ are small compared to $f_{238}$: if both $f_{234}$ and $f_{235}$ are small compared to $f_{235}$, the relative abundances for both $^{234}$U and $^{235}$U simplify to:

$$\frac{f_i}{f_{238}} \approx \frac{f_i}{1-f_i} \tag{S.6}$$

This approximation for $f_{238}$ is most valid at low enrichments for which both the fractions of $^{234}$U and $^{235}$U will be small compared to $^{238}$U. Using 1.4 for the $^{235}$U separation factor, the number of stages can be calculated from Equation S.5. Using this number and a $^{234}$U separation factor of 1.533 (from Wood's approximation), Equation S.5 can easily be rearranged with Equation S.6 to calculate the expected $^{234}$U enrichment. However, as seen in Figure S.1, the fraction of $^{234}$U at high $^{235}$U enrichment is unphysically high, as it is obvious that $1 = f_{238} + f_{235} + f_{234}$ no longer holds above about 83% $^{235}$U enrichment. Above this range, not only must there be no more $^{238}$U in the mixture according to this model, but the sum of the $^{234}$U and $^{235}$U fractions are actually greater than unity.

Assuming $f_{234}$ is small compared to $f_{235}$ and $f_{238}$: if $f_{234}$ is always small compared to both $f_{235}$ and $f_{238}$, all of the relative abundances can be simplified to:

$$\frac{f_i}{f_{238}} \approx \frac{f_i}{1-f_{235}} \tag{S.7}$$

Using the same method and separation factors as the previous model produces a new calculation for $^{234}$U enrichment. As shown in Figure S.1, this result is more reasonable and only becomes unphysical at very high enrichments above 90% $^{235}$U: at 95% $^{235}$U, the $^{234}$U enrichment is 6%. Ultimately, this model does get within an order of magnitude of the available data with only very simple calculations and assumptions.

Constrained by $1 = f_{238} + f_{235} + f_{234}$: the previous model's assumption that $f_{234}$ is always small compared to both $f_{235}$ and $f_{238}$ cannot possibly hold at high enrichments if $f_{234}$ surpasses $f_{238}$, suggesting room for improvement. For instance, we can force the constraint $1 = f_{238} + f_{235} + f_{234}$ to hold. Mathematically, however, this is much more challenging, since now the number of stages in Equation S.5, which depends on $f_{238}$, also depends on $f_{234}$. If Equation S.5 is combined for both $^{234}$U and $^{235}$U, the equations can be rearranged with the full constraint $1 = f_{238} + f_{235} + f_{234}$ to get the following relationship:

$$f_{234} \times (1 - f_{235} - f_{234})^{1-\frac{\ln(\alpha_{234})}{\ln(\alpha_{235})}} = (f_{235})^{\frac{\ln(\alpha_{234})}{\ln(\alpha_{235})}} \times \frac{\left({f_{234}}/{f_{238}}\right)_{natU}}{\left({f_{235}}/{f_{238}}\right)_{natU}^{\frac{\ln(\alpha_{234})}{\ln(\alpha_{235})}}} \qquad (S.8)$$

This equation can be solved using one of the built-in numerical solvers (vpasolve) in MATLAB for the desired range of $^{235}$U enrichments. This result, shown in Figure S.1, is not all that different from the previous model. Notably, the data cuts off at close to 90% enrichment because the solver could no longer find any real roots, suggesting there is a deeper issue with the model, likely in the assumptions about the separation factors. Based on Figure S.1, this and the previous model provide decent results up until about 80% $^{235}$U enrichment.

Semi-Empirical Model: while the previous two models stay within an order of magnitude of the available data, improvements must still be made in order to predict damage to centrifuge materials at high (weapons-grade) $^{235}$U enrichments. Empirical data suggest that at about 93% enrichment, the fraction of $^{234}$U is actually closer to 1% *(44, 45)*, where our models so far predict around 5-6%. Simulations using the MSTAR or M* software or modifications thereof seem to be in closer agreement *(46, 61, 64)* with this value of 1% as well. To bring our model into alignment with the prior literature, we take a semi-empirical approach.

By combining Equation S.5 for both $^{234}$U and $^{235}$U enrichment for the case of natural uranium input and 93% $^{235}$U enrichment with 1% $^{234}$U enrichment output, a separation factor of $\alpha_{234} = 1.42$ for $^{234}$U can be obtained, which is much lower than either Von Halle or Wood predict. Using the more complicated model described in Equation S.8 along with this empirically-informed $^{234}$U separation factor, we produce the semi-empirical model shown in Figure S.1. This treatment does indeed provide a more conservative estimate of the fluence, and the results of this semi-empirical model agree well with the aforementioned data from literature.

Based on this model, the calculation of fluxes and doses are as follows. The majority of the radioactivity in a centrifuge, and therefore the alpha flux causing radiation damage, has now been shown to originate from $^{234}$U. With the relative amounts of each isotope determined as a function of $^{235}$U enrichment, a flux of alpha particles can be calculated in the following manner. We first calculate the density of the $UF_6$ gas using some assumptions about the characteristics of the centrifuge or pipe containing it. Assuming an infinite cylindrical pipe with a 5 cm inner diameter ($d$) at 50 Torr ($P$) and 300 K ($T$) *(66)*, the density of the gas can be approximated using the ideal gas law $\rho = RT/P = 2.67 \times 10^{23}$ mol/cm$^3$. The total number density of the gas is therefore $N_{tot} = \rho \times 6.02 \times 10^{23} = 1.61 \times 10^{18}$ molecules/cm$^3$.

Each isotope will have a volume-specific activity of $A_i = N_i\lambda_i$ based on its decay constant $\lambda$, related to the half-lives in Table 1 by $\lambda = \ln(2)/t_{1/2}$. Here $f_i$ is the atomic (and molecular) fraction of each isotope, and in a three-component system, $1 = f_{238}+f_{235}+f_{234}$. Thus $N_i = f_i \times N_{tot}$. If we now consider a segment of pipe of length $L$, and assuming all the alpha particles decaying in this volume hit the wall surrounding it, the flux $\Phi_i$ in α/cm$^2$yr from each isotope is:

$$\Phi_i = \frac{d}{4}\lambda_i N_{tot} f_i \qquad (S.9)$$

and the total alpha particle flux is the sum of the component fluxes.

For the calculation of the corresponding doses, the contributions from $^{235}$U and $^{238}$U are neglected due to their very small contribution to the total flux as shown in Figure 1 of the main paper. Using the SRIM software package in quick Kinchin-Pease mode and a PTFE composition of $C_2F_4$ with a density of 2.2 g/cm$^2$, an alpha particle range of 21.3 $\mu m$ is obtained using 4.76 MeV alpha particles from $^{234}$U. The dose incurred by the PTFE being irradiated is as follows, assuming that all the alpha particle energy is deposited into that thin volume:

$$dose\ \left[\frac{kGy}{yr}\right] = \frac{4.76\ \left[\frac{MeV}{\alpha}\right] \times 1.6 \times 10^{-13} \left[\frac{J}{MeV}\right]}{2.2 \left[\frac{g}{cm^3}\right] \times 21.3 \times 10^{-4}\ [cm]} \times \Phi_{234} \left[\frac{\alpha}{cm^2 yr}\right] \qquad (S.10)$$

SRIM Simulations and Damage Profiles

Damage profiles were simulated in SRIM-2013, using the ”Quick Kinchin-Pease” damage calculation and the built-in compound correction factor for Teflon (ICRU227) which was 0.955 or −4.52%. The output for 10,000 4.5 MeV He ions is shown in Figure S.2 as the (A) ionization and (B) vacancy production profiles. The longitudinal range was 19.5 $\mu$m with a straggle of 0.3 $\mu$m. For ease of comparison to the FSC data, this was rounded to an approximate range of 20 $\mu$m for 4.5 MeV $He^{2+}$ ions in PTFE.

Comparison of radiation profiles is as follows. While the SRIM simulation is appropriate for predicting the effects of the accelerator-produced ions in our study, the more complicated $UF_6$ gas would produce a different profile in the surrounding material. Natural radioactive decay emits isotropically, so that radiation incident on the surrounding surfaces would be from many directions. The $UF_6$ gas will also attenuate some of the radiation emitted, resulting in a range of energies, less than or equal to the decay energy, incident on the surface. With the combination of these two effects, rather than a classical Bragg peak as in the SRIM simulations, the overall energy loss as a function of depth will exhibit a strictly decreasing behavior. Only the radiation emitted closest to and in the direction of the surface would be able to make it the total possible range (about 20 $\mu$m for 4.5 MeV alpha particles). his is illustrated by GEANT4 simulations, as shown in S.2C, in which for three different uranium enrichments, the energy deposited as a function of depth in the PTFE is shown (as partitioned by source isotope).

The GEANT4 simulations modeled a 5 cm diameter, 5 cm length cylinder of $UF_6$ gas with molecular density $1.6 \times 10^{18}$molecules/cm$^3$ at 5%, 20%, and 90% $^{235}$U enrichments. Corresponding $^{234}$U enrichments were estimated to be 0.04%, 0.18%, and 0.95%. The Shielding physics list was employed, as it includes highly accurate ion interaction models (quantum molecular dynamics) and is recommended for shielding applications *(67)*. 100 million alpha particles for each uranium decay energy at each enrichment composition were simulated, isotropic in the cylinder volume. The PTFE was modeled as 50 cylindrical shells around the cylinder with 1 $\mu$m thickness each. To avoid any edge effects, the result is based on only central 50 $\mu$m section of the modeled PTFE shells. The energy deposited per $\mu$m-PTFE bin is reported in Figure S.2C. The y-axis is reported in MeV/$\mu$m/s because each contribution is weighted by the volumetric activity (in Bq) of that component, since each component is the result from 100 million simulated particles.

DSC Settings and Parameters

Conventional DSC experiments were performed on a TA Instruments Discovery DSC. The base DSC settings were left as defaults, including the cover gas (99.999% pure nitrogen) flow rate of 50 mL/mm and a sampling rate of 0.1 s/pt. The reference pans used in experiments were the TZero aluminum reference pans with matching lids. Each pan and lid combination was measured on a Mettler-Toledo microbalance with a resolution of 0.0001 mg. The pans had average combined masses of 49.9396 mg, while the masses for all the sample pan and lids can be found in the ”Sample masses.xlsx” spreadsheet in the Github data repository for this manuscript. The machine was previously calibrated according to standard procedure using a sapphire heat capacity standard, the calibration file for which is also included in the data repository.

Heating and cooling rate optimization study: Figure S.3 shows data from which the parameter settings (heating/cooling rates, temperature region of interest) were chosen. The goal was to maximize the area under the melting and recrystallization peaks (higher heating rates are better), while incurring no temperature shift due to the time it takes heat to move from the DSC heater to the sample (lower heating rates are better). The heating and cooling rates were chosen to be 10 °C/min. For faster rates, the signal and peak are more pronounced, as shown in Figure S.3A. However, lower rates allow more time for reactions to occur without additional thermodynamic effects due to under cooling or overshooting the melting and recrystallization temperatures.

Number and justification of heating/cooling cycles used for analysis: Analysis was conducted on the second heating/cooling cycle for each measurement. The first melting of the sample often showed a marked difference to subsequent cycles due to the differences in thermal history before and after the first melting, poor initial thermal contact with the sample pan before the first melting, or residual stresses in the sample due to skiving during manufacture or other processing. Since the 5 minute isothermal segment between the first heating and cooling erases thermal history, cooling measurements, even from the first cycle, tended to be much more consistent with each other. Consecutive cycles show a slight drift in the peaks, which can cause marked changes to the peak shape as seen in Figure S.3B. Excluding the first heating, differences between the 1st and 2nd cycles from the rest of the cycles are rare, with visible differences mostly in the baseline making the peak areas still very similar and consistent. A quick peak enthalpy analysis using the TRIOS software package in a temperature region of interest from 250-375 °C shows that the recrystallization enthalpies increase slightly with consecutive cycles, gaining about 1 J/g between the second and tenth cooling cycles. However, the drift is small compared to sample-to-sample variation (Figure S.4), meaning that comparison of the same cycle between samples is sufficient. Compared to Figure S.3B, the second and third cycles are quite consistent with each other, and therefore only the second heating/cooling cycle is used in further experiments and analyses.

Effect and Location in PTFE Film on DSC Results and Its Correction

The relatively large variation in the unirradiated control sample measurements, as seen in Figure S.4, was linked to the location of each sample from the PTFE film stock. This effect is likely related to the manufacturing method of the film stock (skived), whereby small variations in skiving thickness may produce differences in peak enthalpies. To quantify and control for this effect, samples were taken in a grid from a section of the stock, as shown in Figure S.5A. The data grouped very well by the columnar location of the samples, as shown by the baseline-subtracted recrystallization peaks in Figure S.5B. Figures S.5C and d show that these measurements when converted to enthalpies for both the heating and cooling data correlate strongly with the x (columns) dimension. Using this finding, the samples for the main study were taken from a narrow column of the stock material, spanning less than 4 cm. There is a much less obvious correlation with location for the unirradiated control samples in the main study (Figures S.5E and F), as well as a smaller standard deviation for the group of control samples compared to the previous study.

FSC Settings and Parameters

Fast scanning calorimetry experiments were performed on a Mettler Toledo Flash DSC 1 in 99.999% pure argon with a flow rate of 50 mL/min. The heating and cooling rates were chosen as 500 K/s with a sampling rate of 10,000 pts/s and 0.1 s isothermal segments between each heating and cooling ramp. Samples were placed on the Flash DSC UFS1 chip sensor with an eyelash via static attraction, the preferred method for sample placement according to the company. To reuse chips, samples were carefully pried off and removed with the same eyelash. 104 individual samples were measured on a total of 11 different chips. Each chip's calibration information was loaded from Mettler Toledo's pre-compiled database. Before use, the chips were conditioned using the default conditioning method six times, involving multi-second high temperature annealing to remove chip film stresses, after which the default correction method was run once. Most samples were measured multiple times due to issues with thermal contact.

FSC Heating and Cooling Rate Study: A quick cooling rate study was performed on an unirradiated, microtome-sectioned test sample between 1 and 500 K/s to test lower heating rates, from which 500 K/s offered the best signal to noise ratio, as shown in Figure S.6. In order to ensure that the samples had enough time to melt during the dynamic measurement, a heating rate of 500 K/s was chosen. A slower cooling rate was not chosen so that the samples would experience less quenching during recrystallization *(58)*.

Determination of the Number of Cycles for Each Study: Each sample measurement had four heating/cooling cycles. Because the melting peaks were less easily defined, sometimes with multiple peaks (a small example of this is visible in Figure S.11A), analysis focused on the recrystallization peaks. Generally, the third and fourth cycles had converged, so the third cooling was used for the analysis. However, because there were issues with ensuring thermal contact between the samples and the sensor, the measurements in the main FSC study do not necessarily compare the samples after the exact same conditions, since some have been heated more than others. This effect is probably not too great, as the samples are only held at high temperatures for 0.1 s between heating and cooling ramps.

<u>DSC and FSC Data Processing</u>

The data processing for both the conventional DSC and FSC experiments was similar, with some small differences, as shown in Figure S.7. While both methods use spline baselines to isolate the peaks, the parameters to calculate the splines are different due to the slight differences in curve shapes between the two methods. The spline baseline line was chosen because it isolated the effects of irradiation better, as shown by the lower relative deviation for the spline baseline in Figure S.7C. All data were processed in MATLAB, with scripts and original data files available in the GitHub repository for this manuscript.

DSC Data Processing: Data for the conventional DSC experiments were processed in MATLAB using a series of scripts, mainly Heat2_Apr19_finalAnalysis.m for the heating data and Cool2_Apr19_finalAnalysis.m for the cooling data. The original TA Instruments software TRIOS was used to export the data to Excel, and then again to .CSV format to load into MATLAB using the custom auto-generated function importDataCSV(). For ease of data processing, each set of data is truncated between 260 and 360°C and the temperature values adjusted to be evenly spaced, while the corresponding heat flow values are run through the function Data_interp() (linear interpolation) to adjust those values. This results in the data for each sample having easily comparable temperature values, as well as identical array sizes for data handling and baseline subtraction purposes.

The analysis of peak areas (or enthalpies of the melting and recrystallization processes) required construction of spline-based baselines, which could then be subtracted from the data to isolate the peaks. The spline baseline construction is shown in Figures S.7A and B. First, the minimum/maximum of each peak is found and the range of data around it isolated. For the heating curves, this range is -30 to +15°C from the peak temperature, while for the cooling curves it is -20 to +15◦C, to account for asymmetry in the peaks. These ranges are further subdivided to get the outer half of each of these range definitions, which accounts for the straight portions of the data surrounding the peaks. These straight portions are fit to straight lines, and the built-in spline() function is used to create the baseline using the points and slopes of these straight lines to construct continuous splines.

Several variations of baselines were also explored, but the spline baseline resulted in the smallest relative variation in the peak areas calculated. This is shown in Figure S.7C, comparing the results from a linear baseline created from the data endpoints at 260 and 360˚C and the spline baseline as described above. The relative error shown is the standard deviation of each set of measurements divided by the average.

Once the peaks are isolated by subtracting the baselines, the areas are integrated using the built-in trapz() function. This area is converted to an enthalpy by dividing by the heating/cooling rate of 10°C/min. For reporting, the results for each level of fluence are averaged together and the standard deviations calculated.

Correlation to Enrichment and p-value Analysis: The enthalpies measured by DSC are correlated to enrichment in two steps to predict expected enthalpies at key enrichment levels. First, enrichment is correlated to fluence using the semi-empirical model for $^{234}U$ enrichment and shown in Figure 1 of the main paper. Based on this model, the yearly fluence expected from any enrichment can be calculated using Equation S.8 and Equation 1 from the main paper. Similarly, based on the DSC enthalpy measurements shown in Figure S.8, expected mean enthalpies and approximate standard deviations can be estimated. By combining these two estimates, an expected enthalpy with standard deviation can be correlated to exposure for 1 year of a given enrichment. Having performed this extrapolation at enrichments of interest (the red markers in Figure S.8, the extrapolated means and standard deviations can be quantitatively compared to each other, as an assessment of how successful a measurement like this would be at discriminating between different enrichments.

Because all of the fluences of interest (5%, 20%, and 90%) predicted from the model fall between $10^9$ and $10^{11}$, we fit a quadratic polynomial to only the irradiated data (excluding the unirradiated controls). The data was fit to the individually calculated enthalpies for each of the 48 irradiated samples, rather than the mean values displayed in our results. For a fit equation y(x) = p1 × x2 + p2 × x + p3, where x is the base-10 logarithm of the fluence, the coefficients are p1 = 1.483, p2 = −24.71, and p3 = 121.6, which is shown overlaid on our result in Figure S.8. The red dotted lines are the 95% confidence, non-simultaneous, observation prediction bounds calculated from the fit model through MATLAB. For a new measurement at a specific fluence, the fit model predicts that the measurement will fall within the prediction bounds with a 95% confidence, including an uncertainty for normally distributed random error. The fit model has good agreement with the means for each set of fluences, and the 95% prediction bounds match the 2 standard deviation error bars quite well. Between $10^8$ and $10^9$, the model predicts a minimum, which does not match the expected trends if the unirradiated controls were included in the model. However, because we were not expecting to be able to distinguish measurements in that range from the unirradiated controls, the model was accepted anyway. Based on this quadratic function, if we were to take another 12 samples at a fluence between $10^8$ and $10^{11}$, the expected mean can now be obtained from the quadratic function. The standard deviation for such a measurement is estimated from the 68% prediction bounds at this fluence. The predictions for natural, 5%, 20%, and 90% enriched uranium are shown as the red symbols in Figure S.8.

To make a quantitative comparison, we calculate the p-values between ”measurements” of different enrichments using an unpaired two-sample t test (Equation S.11), assuming unequal variances. The sample numbers used in the calculation are 20 for ”unexposed” samples (to match the actual sample number from the set of unirradiated control samples), and 12 for the rest of the points, as this was the sample number for the measured data at each fluence. To get the p-value from the t-statistic, it is necessary to estimate the degrees of freedom, which we do using two different methods. The first is calculated from the Satterthwaite approximation (Equation S.12, which is the built-in method in MATLAB’s ttest2() function, while the second is a much more conservative estimate as the smaller of the sample numbers minus 1. In all cases, this more conservative estimate of the degrees of freedom is therefore 11. In Table S.2, we calculate the p-values, including the results from both degrees of freedom estimates. The two-tail p-values were obtained with the MATLAB function tcdf(), given the calculated t-statistics and degrees of freedom. In both calculations of the p-values, all of the p-values are well below a significance level of 0.01, except for the comparison of a natural enrichment to an unexposed case. In this case, the p-value is about 0.63, meaning that there is a very high probability that these means are not significantly different from each other.

$$t = \frac{x_{a,m} - x_{b,m}}{\sqrt{\frac{s_a^2}{N_a} + \frac{s_b^2}{N_b}}} \qquad (S.11)$$

$$DOF_s = \frac{\left(\frac{s_a^2}{N_a} + \frac{s_b^2}{N_b}\right)^2}{\frac{1}{N_a - 1}\left(\frac{s_a^2}{N_a}\right)^2 + \frac{1}{N_b - 1}\left(\frac{s_b^2}{N_b}\right)^2} \qquad (S.12)$$

Summary of Full Results with Heating Data: The full result for both the heating and cooling data is shown in Figure S.9. The heating data actually shows slightly better sensitivity of peak enthalpy to fluence, but the cooling data is used in the main study due to its better comparison to the FSC results.

FSC Data Processing

The Mettler Toledo Flash DSC 1 uses the software STARe for data collection. The temperature used is the reference temperature or the recorded temperature for the reference side of the Flash DSC chips. This is by convention, and because some transitions or transformations may affect the temperature of the sample side of the chip. Presumably because the Flash DSC is a high heating rate, power-compensation type DSC, the data is collected at highly regular and evenly spaced times and temperatures, which technically negates the need to interpolate similar data to compare sets of data at identical temperatures. However, the data was run through the interpolation algorithm anyway, to ensure that temperatures all matched up.

While each sample was put through four cycles of 500 K/s heating and cooling with 0.1 s isotherms in between for each experiment, only the data from the 3rd cooling, 4th heating, and isotherm before the 3rd cooling underwent further analysis for this work. The 3rd cooling was chosen for enthalpy analysis because samples that varied between runs due to removal of irreversible effects or artifacts would typically converge by the third heating. The 4th heating was used in conjunction with the 3rd cooling to calculate the mass of the sample during the 3rd cooling and 4th heating based on the measured total heat capacity, and dividing by the temperature-dependent specific heat capacity of PTFE from literature. The 4th heating was used rather than the 3rd heating, because the 4th heating is the one that represents the measurement of the melting of the structure formed during the 3rd cooling. The isotherm before the 3rd cooling was used to identify outliers, as an indication of quality of contact between the sample and chip before the main 3rd cooling measurement. For the heating and cooling data, the reference temperature and corresponding heat flow out of the sample (exo up) were used for data analysis, while for the isotherm data, the time coordinate and corresponding heat flow were used.

FSC Baseline Construction and Peak Integration: Baselines for the FSC experiments were also spline-type, but defined slightly differently due to the different shapes of the peaks in the raw data. The baseline construction lines were not centered around the peak. Instead, for all the data, the construction lines were simply fit to the temperature ranges of 295-310°C and 340-355°C below and above the peak. This deviation from the DSC data processing was performed to accommodate much larger variations in peak sizes and shapes, as shown in Figure S.7D. Then splines were constructed in the same manner as in the DSC experiments, subtracted from the data, and integrated.

FSC Data Outlier Identification

Outliers were identified based on the isotherm data. Because the instrument switches between a dynamic heating or cooling to a static temperature at the isotherm, it exhibits a controlled response, as the instrument uses temperature feedback to attain an equilibrium state during the isotherm. This is necessary because the sample is not in a perfectly adiabatic environment. Considering a very simple energy conservation, during the dynamic segment, the sample will experience heat storage, heat loss to the environment, and conductive heat transfer with the chip. During the static portion, the temperature remains constant, but the heat flow out of the sample must change once equilibrium is achieved, because change in heat storage through the inherent temperature dependent heat capacity will become negligible at equilibrium. However, since this equilibrium is achieved through a programmed, controlled response, it will exhibit characteristics of the controller, assumed to be of the PID (proportional–integral–derivative) or PI (proportional–integral) type.

In fact, the samples show a range of damping behaviors during this isotherm, which can be characterized using quantitative methods. This damping is attributable to ”disturbances” to the system, based on the quality of interaction between the sample and the Flash DSC chip. For example, a larger sample can store more heat, which in turn would damp the controller’s response more. Similarly, thicker samples would be more difficult to keep at a uniform temperature all the way through; samples that do not lay flat and stick up into the cooler environment would lose more heat to the surroundings; poor thermal contact due to uneven surfaces would make it harder for heat to be conducted from the chip to the sample; and reactions occurring that release or absorb heat would make it more difficult to maintain a temperature. All of these examples would provide a damping effect on the sample with the exception of the last, which could affect the sample either way depending on the type of reaction. For this reason, a sample deemed over-damped can be identified as an outlier and rejected from the data set for having a poor-quality contact with the Flash DSC chip. In Figure S.10A, this principle is illustrated, whereby visible differences in the damping coefficient are evident between samples of different mass and contact quality. As a final note, concerns of poor-quality measurements led to taking repeated measurements of the same sample a few times to achieve a higher quality data point.

Quantitatively, this identification is achieved by fitting the unit step response of the transfer function of a basic second order ordinary differential equation to the isotherm data that has been centered so that the tail starts at 0 and scaled so that the equilibrium achieved lies at 1. While a 2nd order system does not quite capture the intricacies of a PID controller, it can describe its response from the point of view of a simplified, 2nd order linear differential equation of a damped oscillator. By modelling it with this simpler system, it becomes very easy to extract a single quantitative measure of damping. Logistically, this required a two-step fitting process. First, we made a function y = transferFit(x,k,w,z) that takes a series of x coordinates, k the steady state gain of the transfer function, w the natural resonant frequency, and z the damping ratio, and computes the step response y for a transfer function of the form $G = k * w2/(s2 + 2 * z * w * s + w2)$ using the built in MATLAB function step(). Second, we used MATLAB's cftool to generated a function createFit tf() that fits our data to the custom transferFit() function. Thus, we can pass createFit tf() a set of centered and scaled data, and have it extract a damping ratio. It is necessary to set a threshold for acceptable levels of damping. We judged that to design a robust system such as the Flash DSC 1, it was likely that the controller was designed to perform mainly in the regime of critically damped to slightly underdamped. Thus, we set my damping ratio threshold for outlier identification to 1.1, or allowing of slightly overdamped samples, but not too many.

### FSC Sample Mass Calculation

Masses were calculated using the method outlined by Cebe et al. *(57)*, based on the measured total heat capacity of the FSC sample compared to either literature values or specific heat capacities measured in an alternative way. To perform this method, a region far away from the melting and recrystallization temperatures was examined for the 3rd cooling and 4th heating, between 87˚C and 177˚C for this work, as shown in Figure S.11. The symmetry method was employed, by fitting straight lines to these segments of the data, averaging them, and then subtracting the average from each segment of data. This adjustment makes the heating and cooling symmetric around zero, as is expected for the measured heat capacity in a DSC experiment (measured heat capacity should be the same upon heating or cooling). Literature values were taken from Wunderlich and Baur *(60)* for ”highly crystalline” PTFE and converted to the more modern units of J/g˚C. At each tabulated point between 87 and 177˚C from Wunderlich and Baur, the measured total heat capacity from the FSC is divided by the tabulated value. All these mass measurements are averaged to give one single mass value. While our measured samples are likely not completely crystalline, due to the high cooling rate of 500 K/s, these reference values at least allow a quantitative analysis of mass in these FSC experiments.

As Cebe et al. mention, this method may be subject to some systematic errors. However, it is considered a sufficient method to distinguish between variation in masses between samples. This is the most important thing for this mass analysis to do, as it provides confidence in the comparison of measurements of different sample shapes (not perfectly round or between series of slices) and validates microtome sectioning reliability.

### FSC Microtome Settings and Sectioning Technique

A Leica EM UC7 Ultramicrotome was utilized for sectioning PTFE stock material for FSC measurements, along with an LN2-cooled cryostage, set between -25 and 0°C. Glass blades were used, freshly made as close to the start of each set of sections as possible to ensure maximum sharpness. Because the rotary action on this specific microtome was not working properly, the 1 µm sections were made by using the stage adjustment settings to advance the blade 1 µm between each section, rather than the sample advancing with each rotation as it would during normal operation. This was judged to be sufficiently accurate and reliable, since the stage adjustment had a fine control of 0.1 µm. Slices were made manually, using the rotary hand-turned wheel. The sample to be microtomed was aligned by eye, using the manual rotation control for the blade's stage to ensure it was parallel to the sample's surface. This was achieved by setting the blade very close to the surface with the lights on both above and below the sample. Close to the sample, but not quite touching, a faint shadow can be seen on the surface. Best results were obtained when this shadow was as parallel to the blade as possible.

The PTFE 1/32" thick stock was prepared for microtoming by forming a roughly rectangular shape with a razor blade, leaving a small nub protruding. The bottom of the larger rectangle provided a flat surface that seated well in the microtome sample holder and would help to align each sample parallel to the blade. The front of the nub was faced off in the microtome before irradiation. These samples were irradiated to a fluence of 1011 4.5 MeV He2+ ions in the CLASS accelerator. After irradiation, a 250 µm brain punch mounted in a drill press was used to create a small, columnar indentation in the samples. Upon microtoming, the successive slices from this diameter-controlled column would be retrieved for FSC. Because smaller faces tended to make more reliable, uniform slices, the sample's nub was further trimmed down to an approximately trapezoidal geometry around the face from which to be sliced. The smooth angled faces allowed the blade to take hold and initiate each slice. After each slice, a human eyelash in a pin vise (same as FSC sample-chip placement) was used to separate the round sample from the detritus and transfer the sample to a prepared glass slide, marked in a grid to identify samples. The glass slide loaded with samples was sandwiched with another slide, separated by a piece of tape at the edges, to protect samples during transport and storage.

The biggest challenge was how to obtain consistent and sequential slices. One of the problems would be when an uneven defect would form on the surface of the sample. This would cause the blade to catch and glance off, only cutting a portion of the surface and introducing a source of error in the thicknesses and depths of the remaining sequential slices. A second pass of the blade would usually, but not always, cut the remaining slice from the surface. In some cases, multiple such defects could exist, and if located on the pre-indented column for the FSC samples, it could result in only a small sliver of actual sample at a reasonably confident depth. In such cases as these, the situation could sometimes be remedied by advancing the blade an extra micron or two to take a thicker slice and remove the defect(s). In some cases, the slivers were too considered too small to separate from the detritus with the eyelash, and so were not retained as samples for FSC. This accounts for some of the missing data points in the series of measurements. Another issue was that the columnar indent using the brain punch was not always deep enough, which resulted in some of the last samples at the greatest depths not being separate enough to be removed from the detritus. This did not constitute a serious problem, as samples were sectioned to depths considerably deeper than the range of 4.5 MeV alpha particles in PTFE, leaving enough unirradiated area for analysis. Lastly, some samples were lost simply by being accidentally flicked off when trying to pick them up with the tip of the hair, dropping off (very rare) when transported to the glass slide, or lost during placement on or removal from the Flash DSC chip through either of these mechanisms.

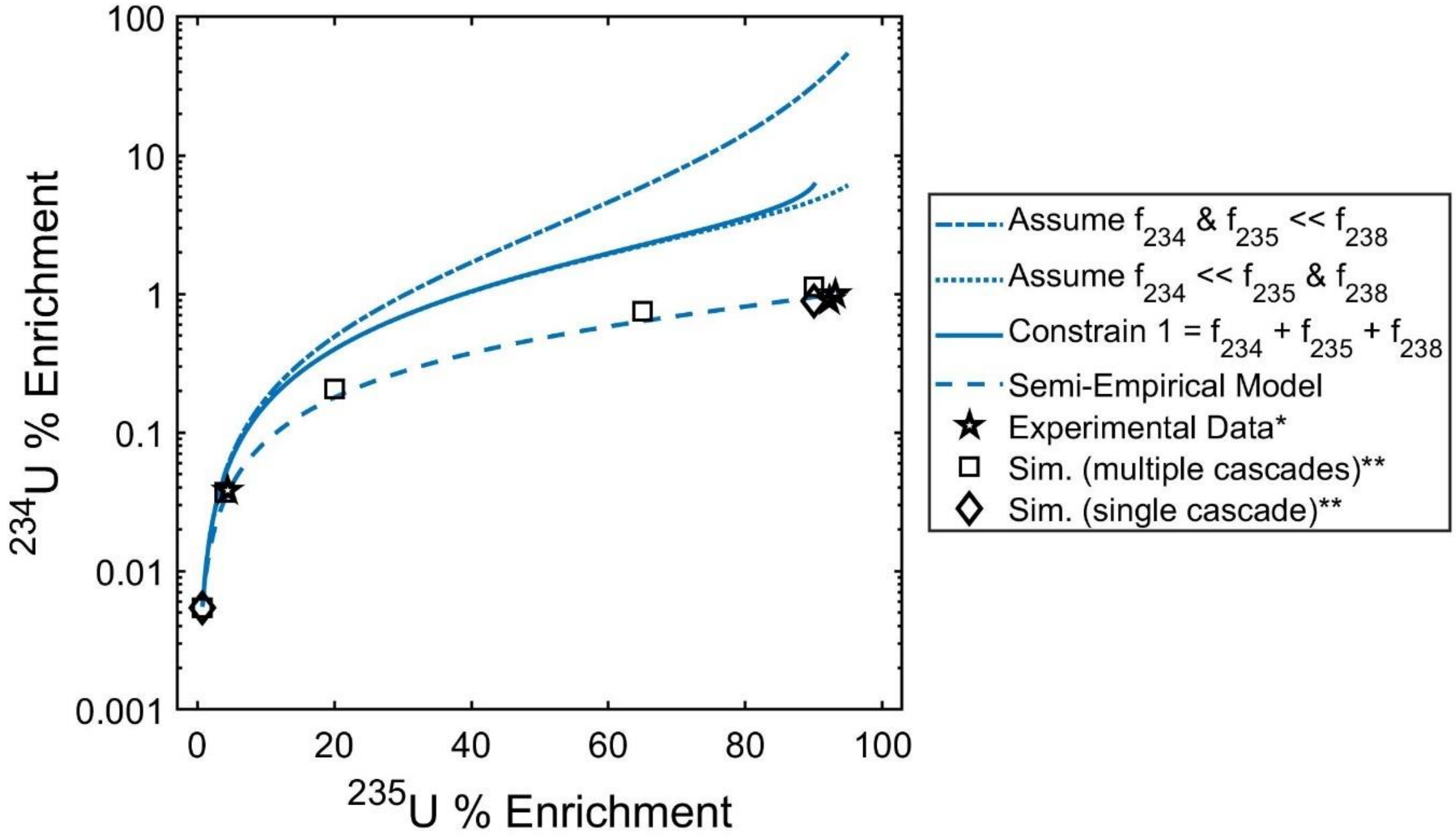

**Fig. S1.**

Comparison of all models considered to estimate the fraction of $^{234}$U based on the amount of $^{235}$U in an enrichment cascade. While all the models are in fairly good agreement at very low $^{235}$U enrichments, at 4% $^{235}$U enrichment the semi-empirical model comes closest to approximating the available data with 0.035% $^{234}$U enrichment. In comparison, the other three models predict 0.05% $^{234}$U enrichment. At higher enrichments, this deviation becomes more pronounced. At 93% $^{235}$U enrichment, which is expected to yield around 1% $^{234}$U enrichment, the closest non-empirical model predicts about 5.4% $^{234}$U enrichment. *Experimental data taken from *(43-45)*. **Simulations results from *(46)*.

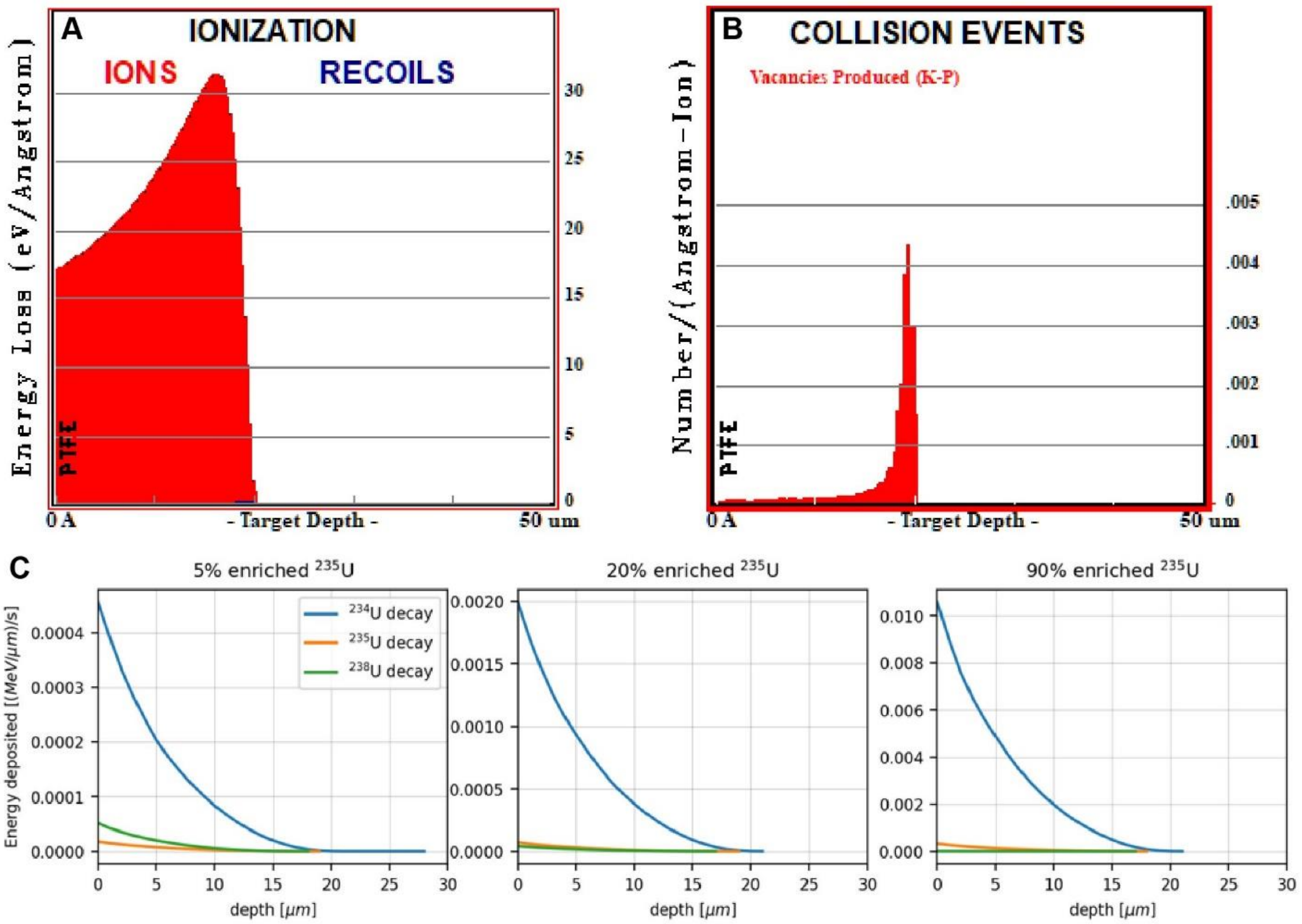

**Fig. S2.**

(A) Ionization profile and (B) vacancy production profile produced in SRIM for 4.5 MeV $He^{2+}$ ions in PTFE. (C) GEANT4-simulated profiles for a cylinder of $UF_6$ gas for comparison.

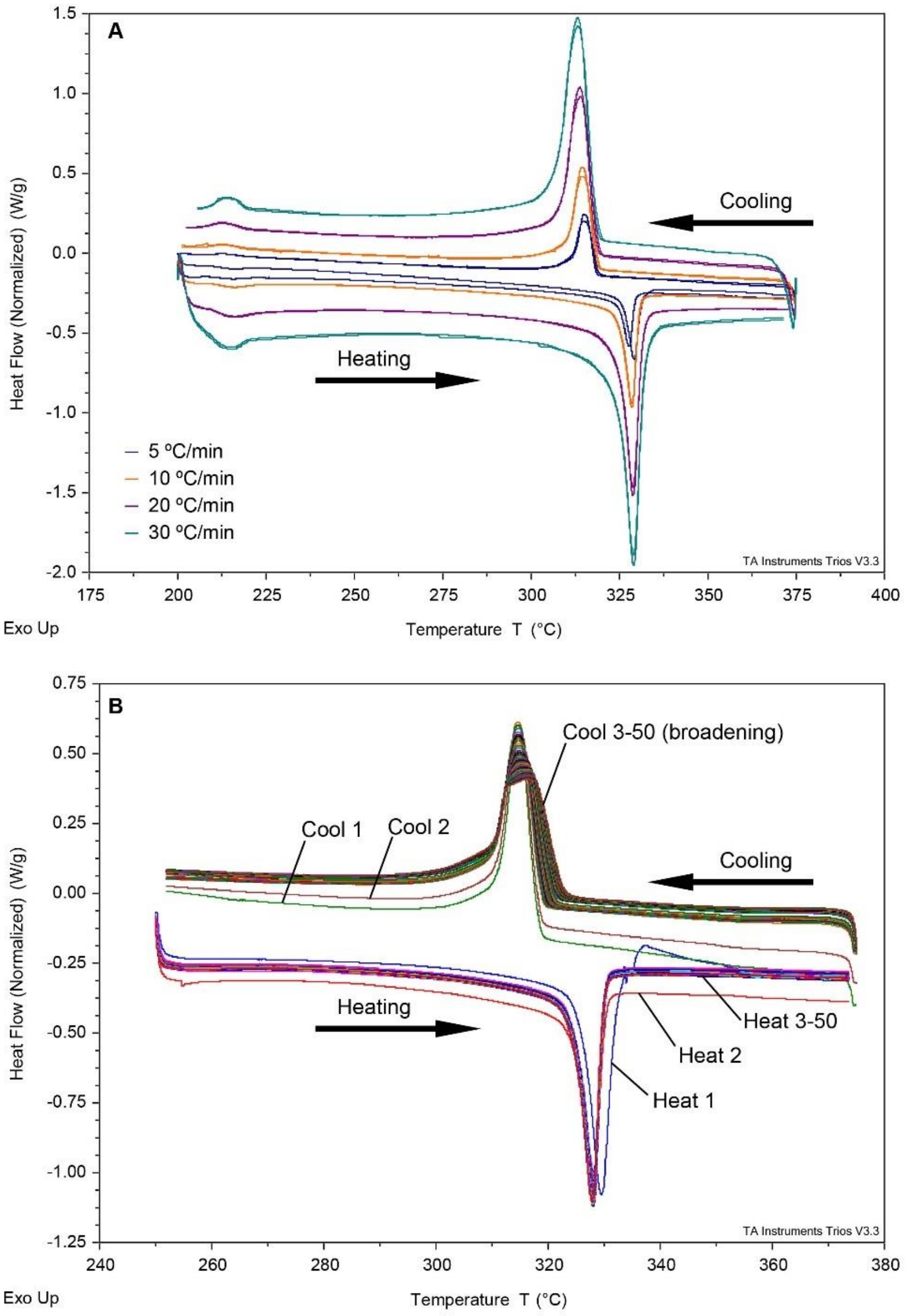

**Fig. S3.**

(A) Effect of changing heating and cooling rates for measurement of 1.4 mg PTFE film. A single sample was used, heated from 200 to 375°C through all four heating rates consecutively before the entire cycle was repeated for a total of eight cycles at four different rates. As the rates increase, the peaks broaden and increase in height, and they experience a slight temperature shift towards lower temperatures on the cooling cycle. The second set of cycles show small decreases in peak height and slight broadening compared to the previous measurements at those rates. (B) Effect of thermal cycling is checked by measuring a 1.4 mg film sample for fifty consecutive cycles, between 250 and 375°C at 10°C/min. In the cooling curves, successive cycles show peak heights diminishing and broadening, while the heating peaks barely change after the third cycle.

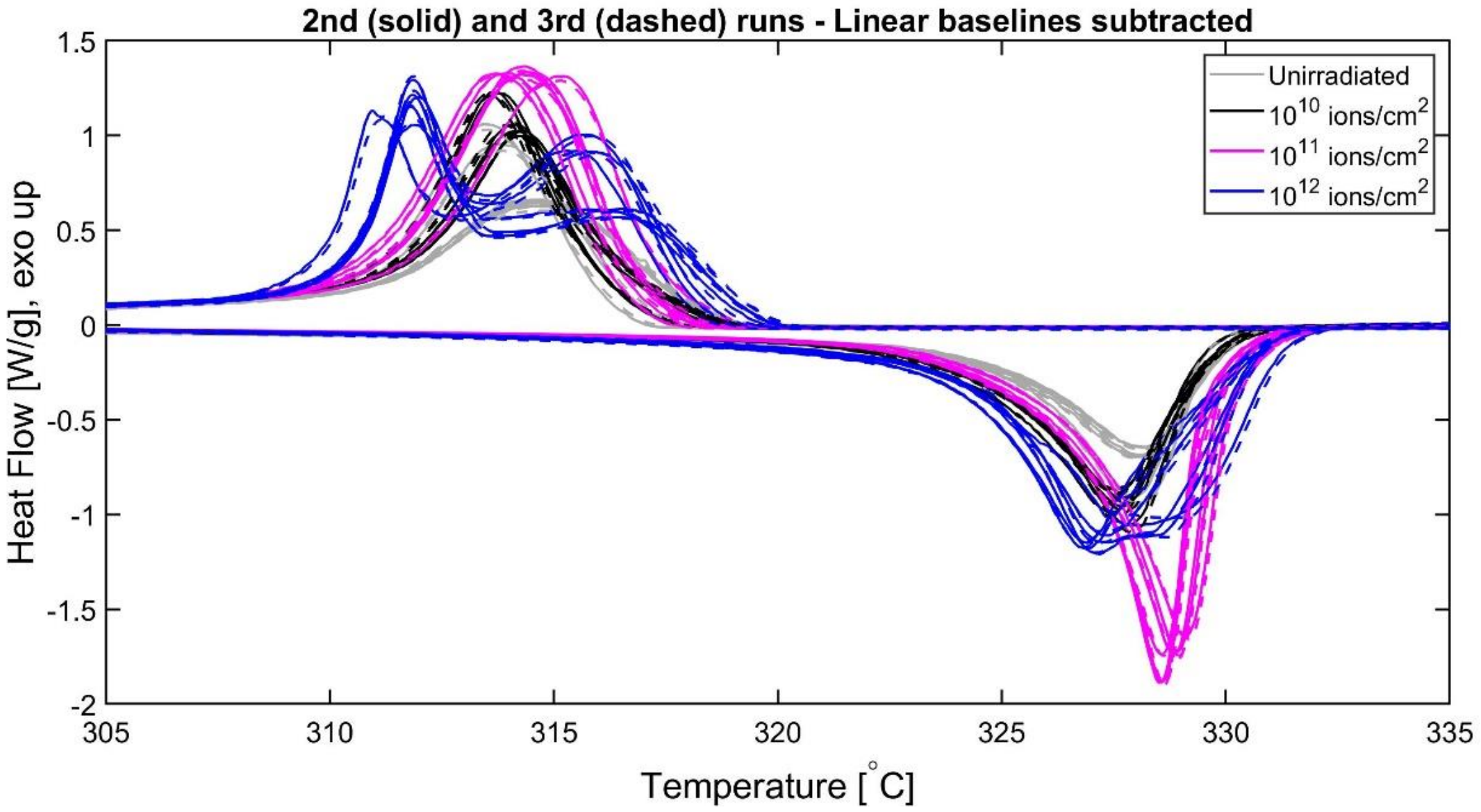

**Fig. S4.**

A previous set of measurements at 10°C/min between 250 and 375°C which includes a third cycle, as well as one more order of magnitude in fluence. Measurements are adjusted for comparison by the subtraction of a linear baseline defined by the data endpoints between 260 and 360°C. Drift of peaks between consecutive cycles is much smaller than the sample-to-sample variation. The double peak behavior at the highest fluence has been observed in literature for different types of irradiation *(36, 68)*.

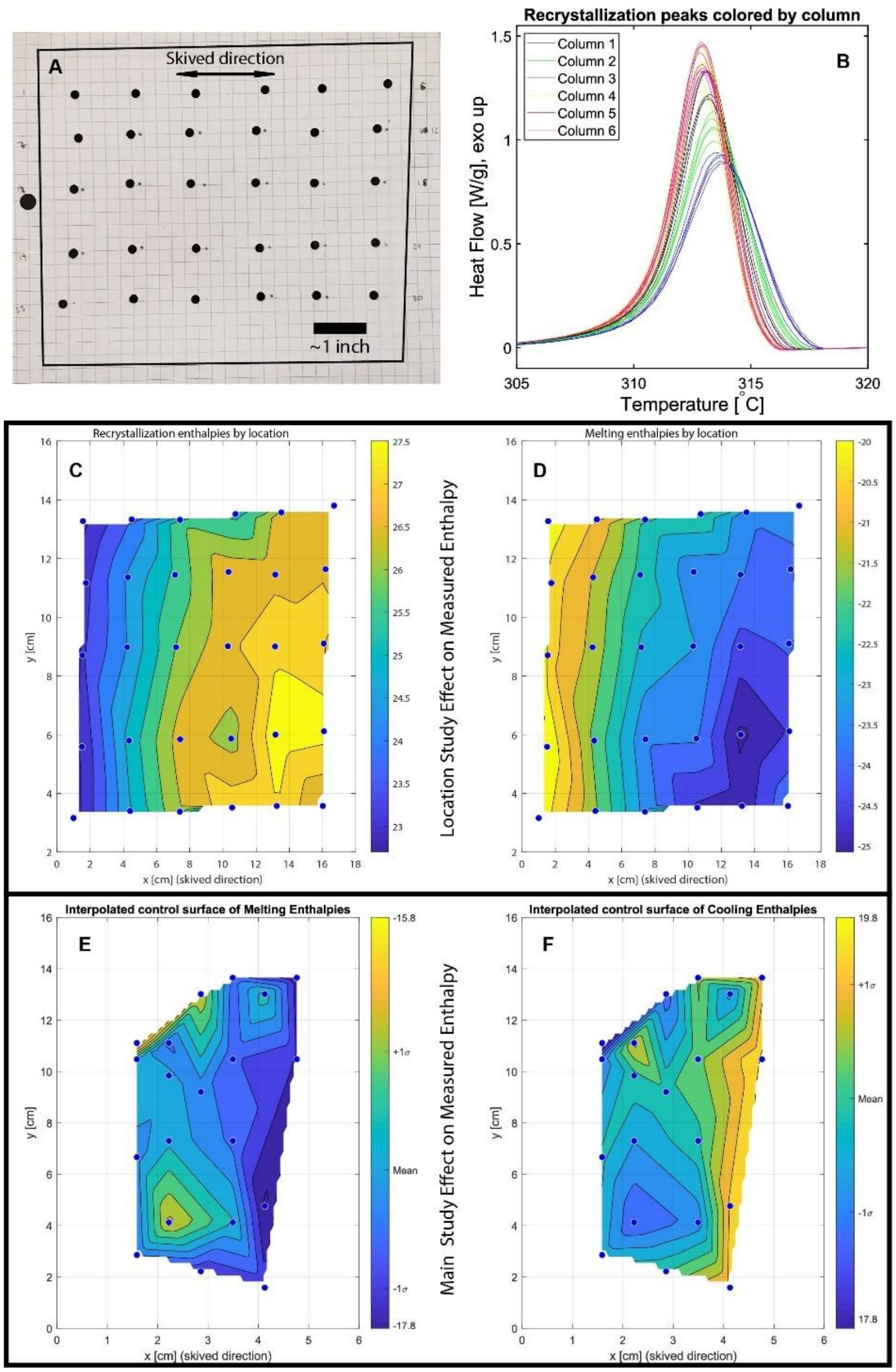

**Fig. S5.**

Elucidation of control enthalpy surfaces from PTFE samples, for the purpose of more accurate control (unirradiated) baseline subtraction from irradiated specimens. (A) Image shows how samples were taken. The film is essentially transparent, so graph paper underneath the film was used to space out the sample matrix. The stock material used was roughly a 6” x 7” rectangle. (B) DSC scans of unirradiated PTFE from each column, showing good grouping of results compared to adjacent columns. (C-D) Calculated control enthalpy surfaces for specimens used in this location-dependent study. (E-F) Calculated control enthalpy surfaces for specimens used in main study.

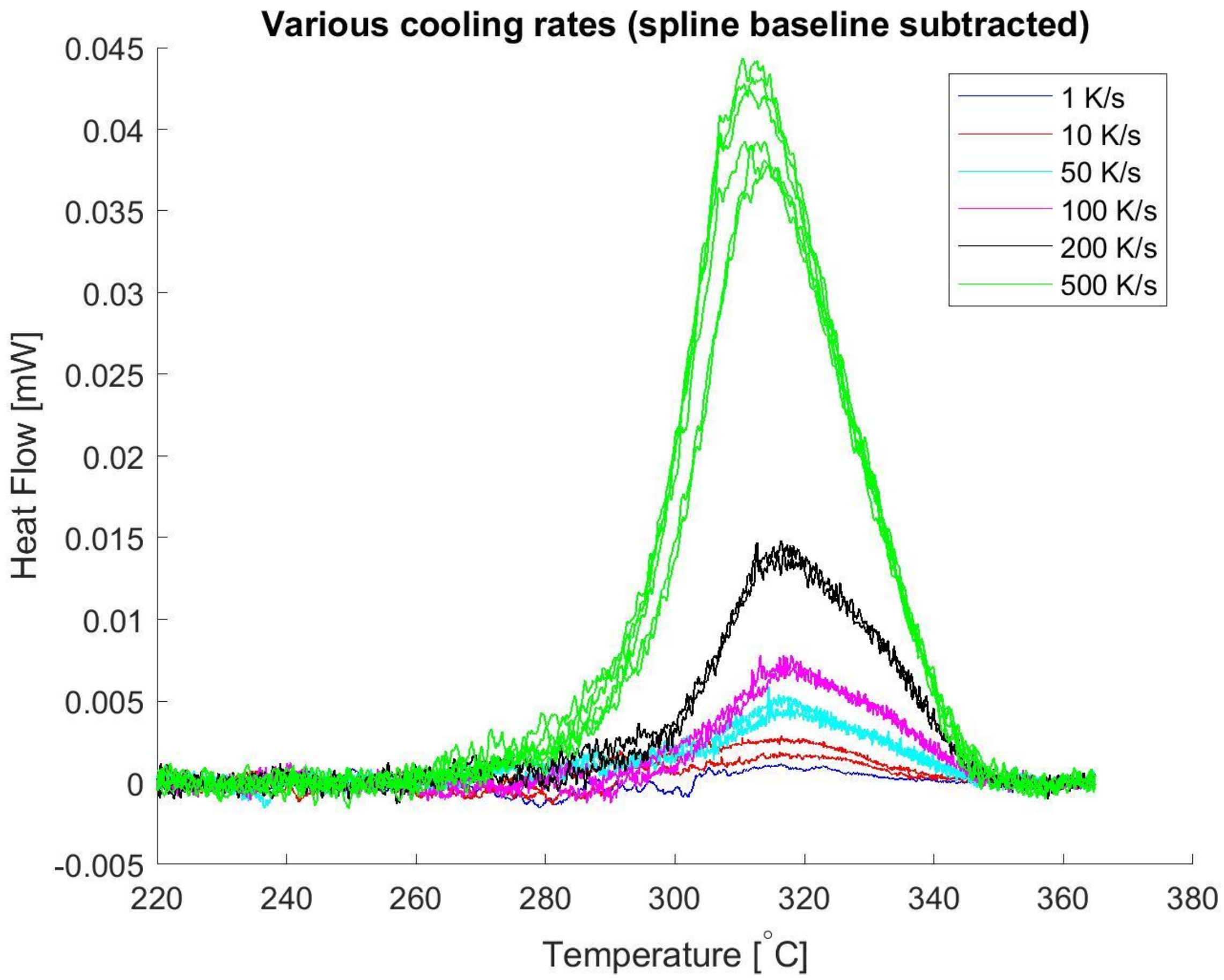

**Fig. S6.**

Measurement of recrystallization of an unirradiated PTFE sample at various cooling rates. Prior to measurements shown, samples were melted at the same rates and held at the maximum temperature for 0.1 s.

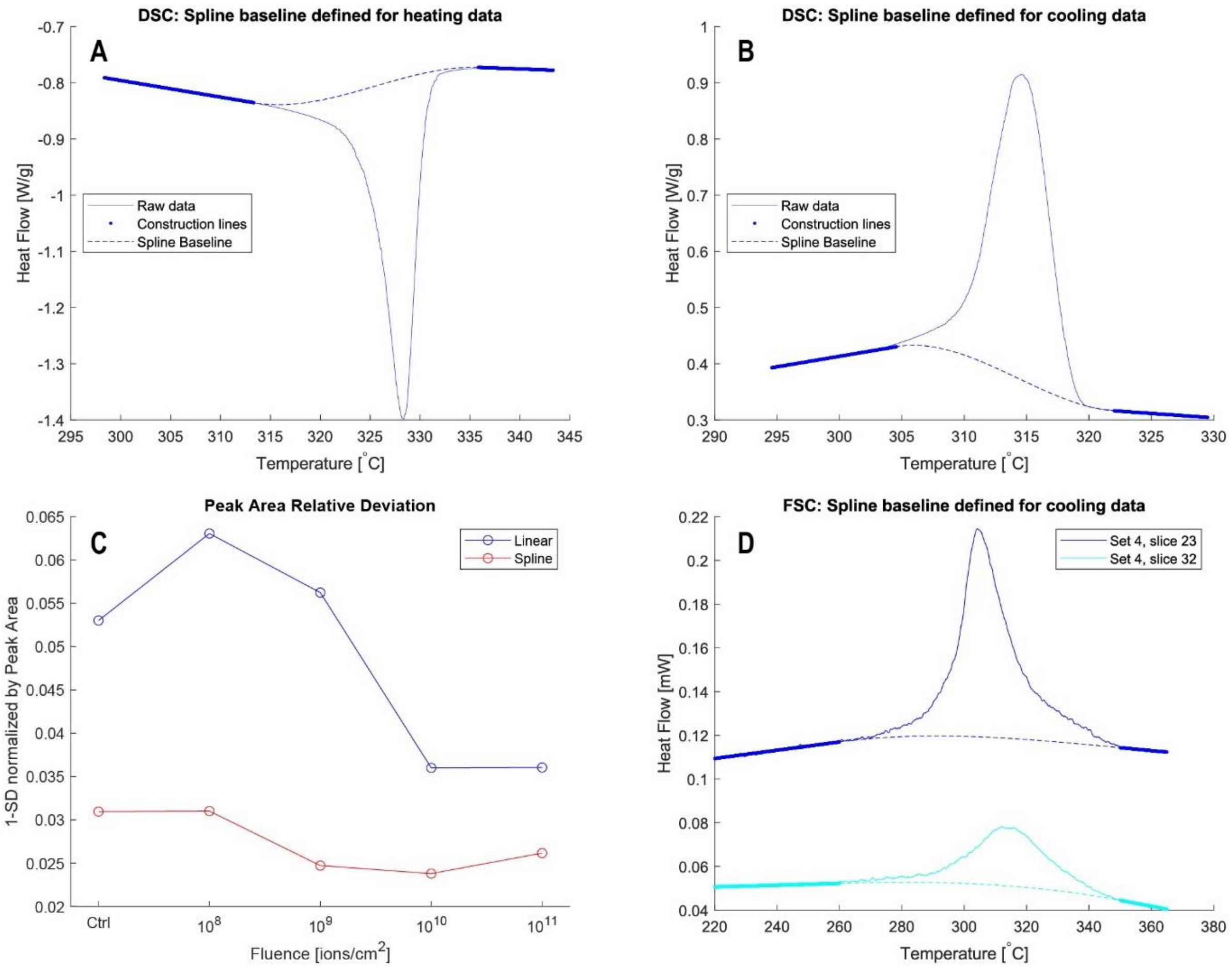

**Fig. S7.**

Baseline optimization study to minimize error between DSC and FSC scans. (A) Spline baseline for melting peak in conventional DSC. (B) spline baseline for recrystallization in conventional DSC. (C) Relative deviations for peak areas calculated using a linear vs. a spline baseline for conventional DSC experiments. (D) Spline baseline for recrystallization in FSC.

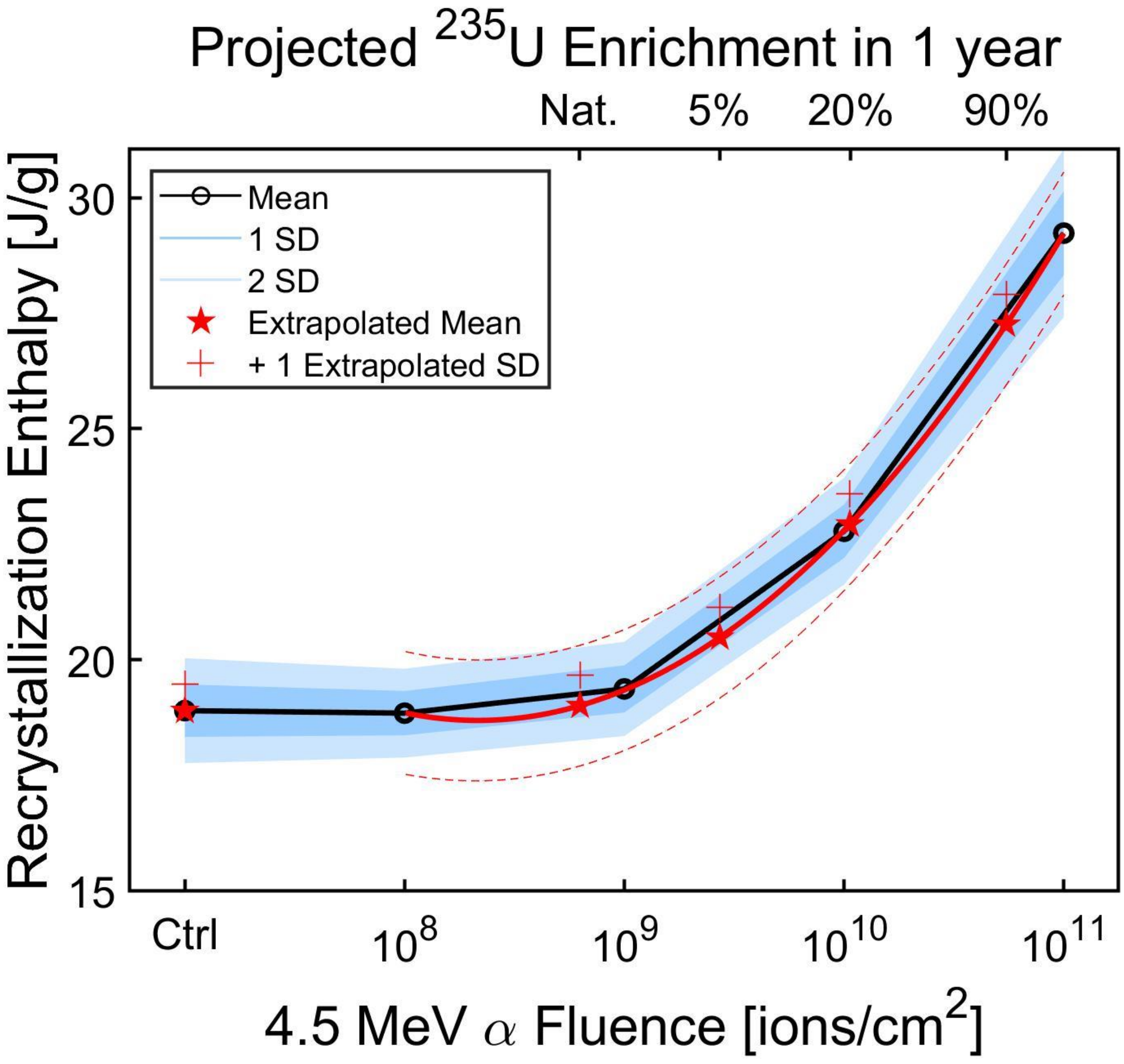

**Fig. S8.**

Expected fluences at unexposed, 0.72% (natural uranium), 5% (typical LEU), 20% (legal limit for HEU), and 90% (weapons-grade) enrichments based on a quadratic fit to only the irradiated data. The quadratic was fit to the measured enthalpies of the 48 irradiated samples. Red cross symbol marks the 1 expected standard deviation, based on the 68% prediction bounds of the quadratic fit.

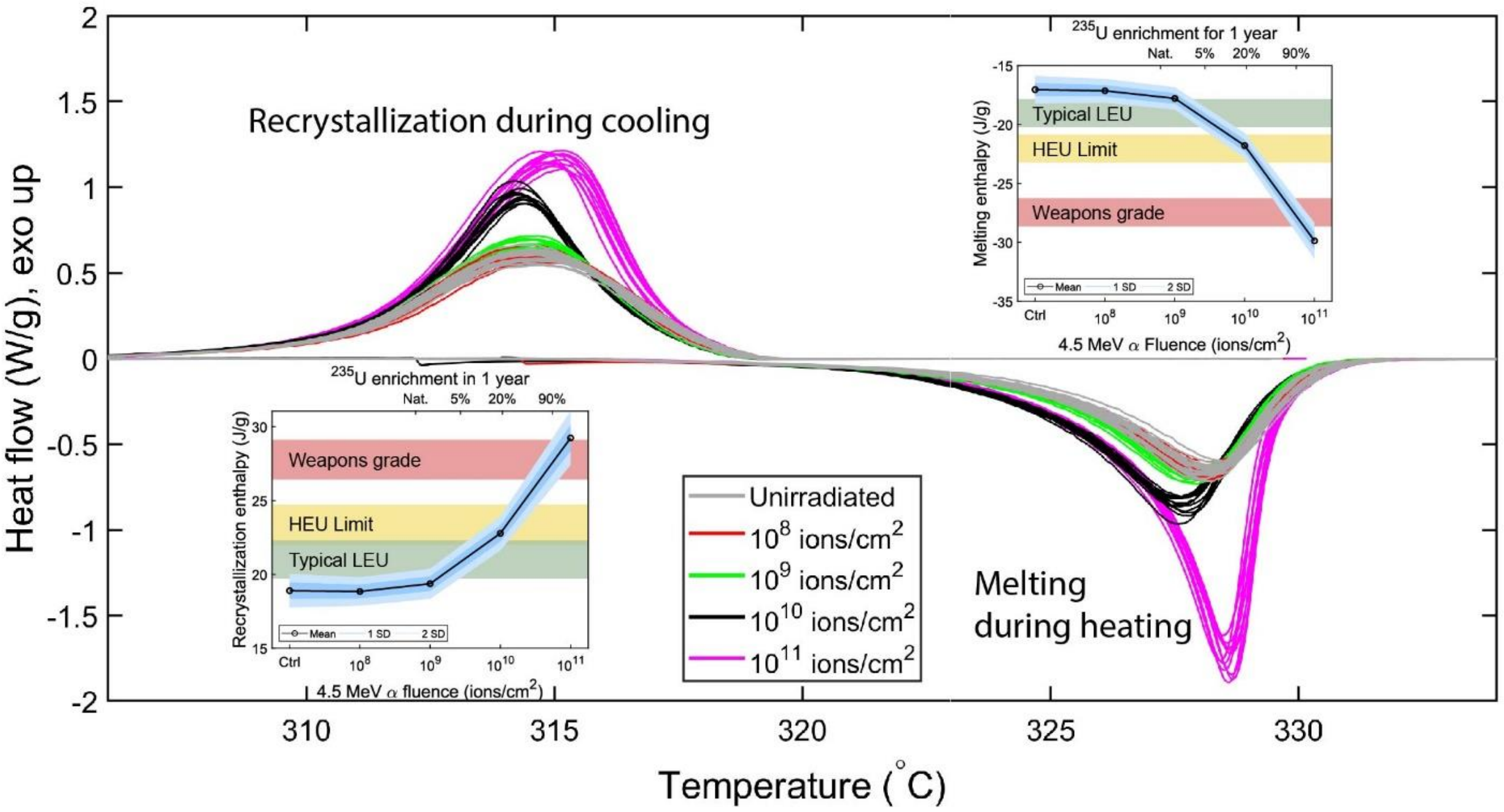

**Fig. S9.**

Baseline corrected DSC data colored by fluence. Insets show the averaged enthalpies for both the melting and recrystallization peaks.

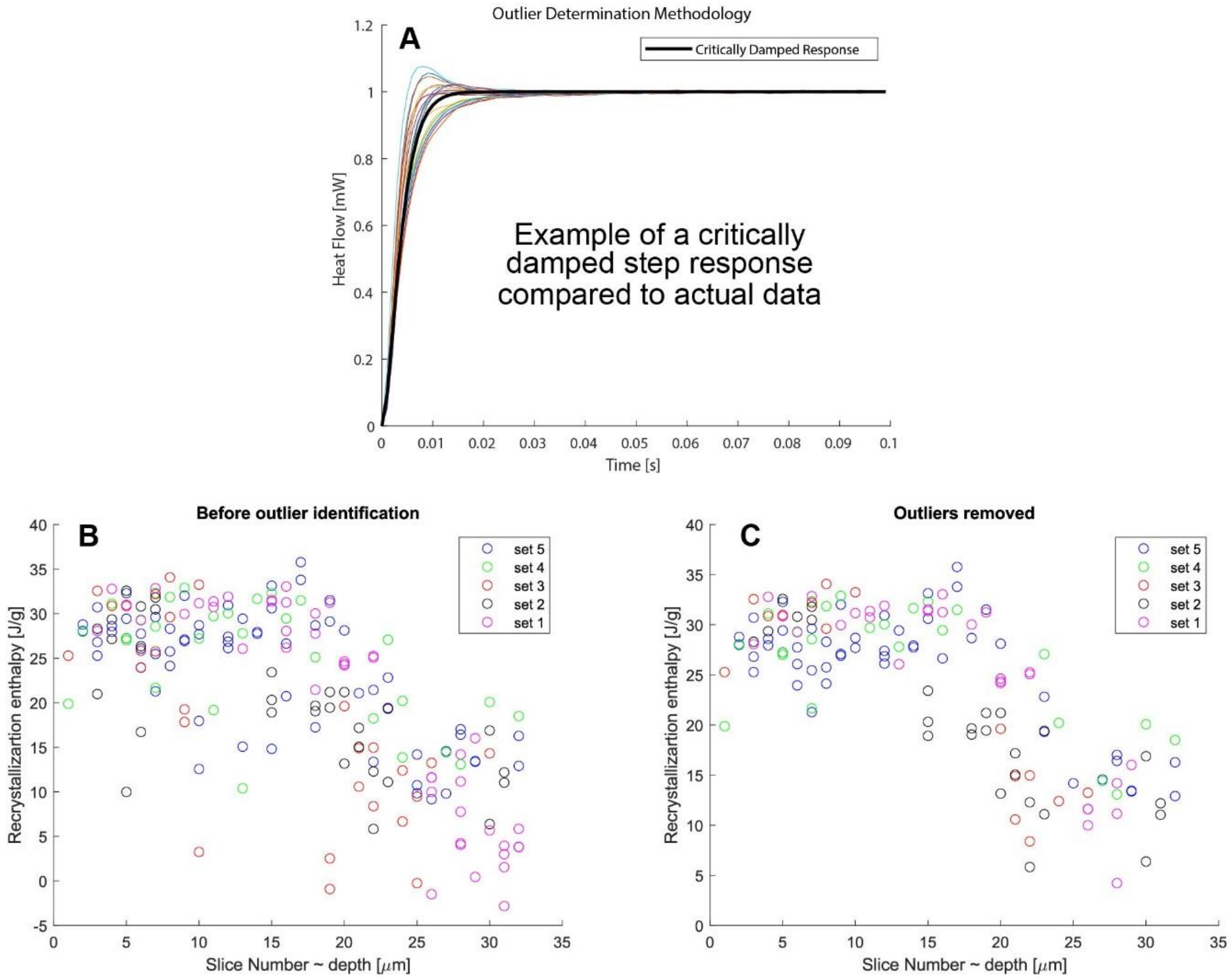

**Fig. S10.**

Identification of outliers in Flash DSC data via signal damping ratio analysis. (A) Isotherms for a subset of the data, with an example of a critical response overlaid. (B) All of the collected FSC enthalpy data, processed without removing outliers. (C) The same data after removing outliers.

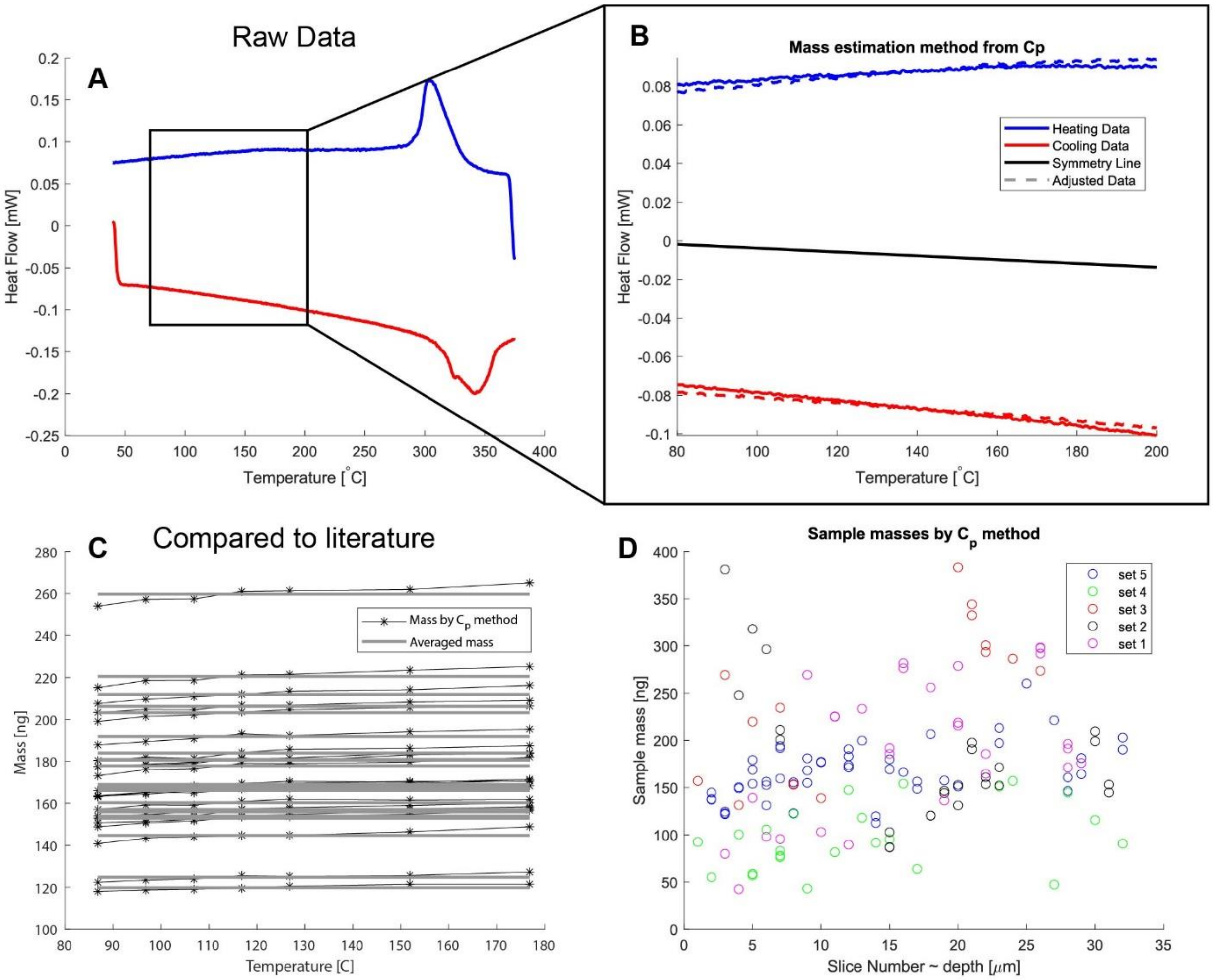

**Fig. S11.**

Determination of FSC sample masses using the total heat capacity method. (A) Example of the raw FSC data. (B) Symmetry correction to the data. (C) Conversion of measured total heat capacity to mass through comparison to literature values of temperature-dependent specific heat capacity. (D) Measured masses of all the measurements, after removal of outliers.

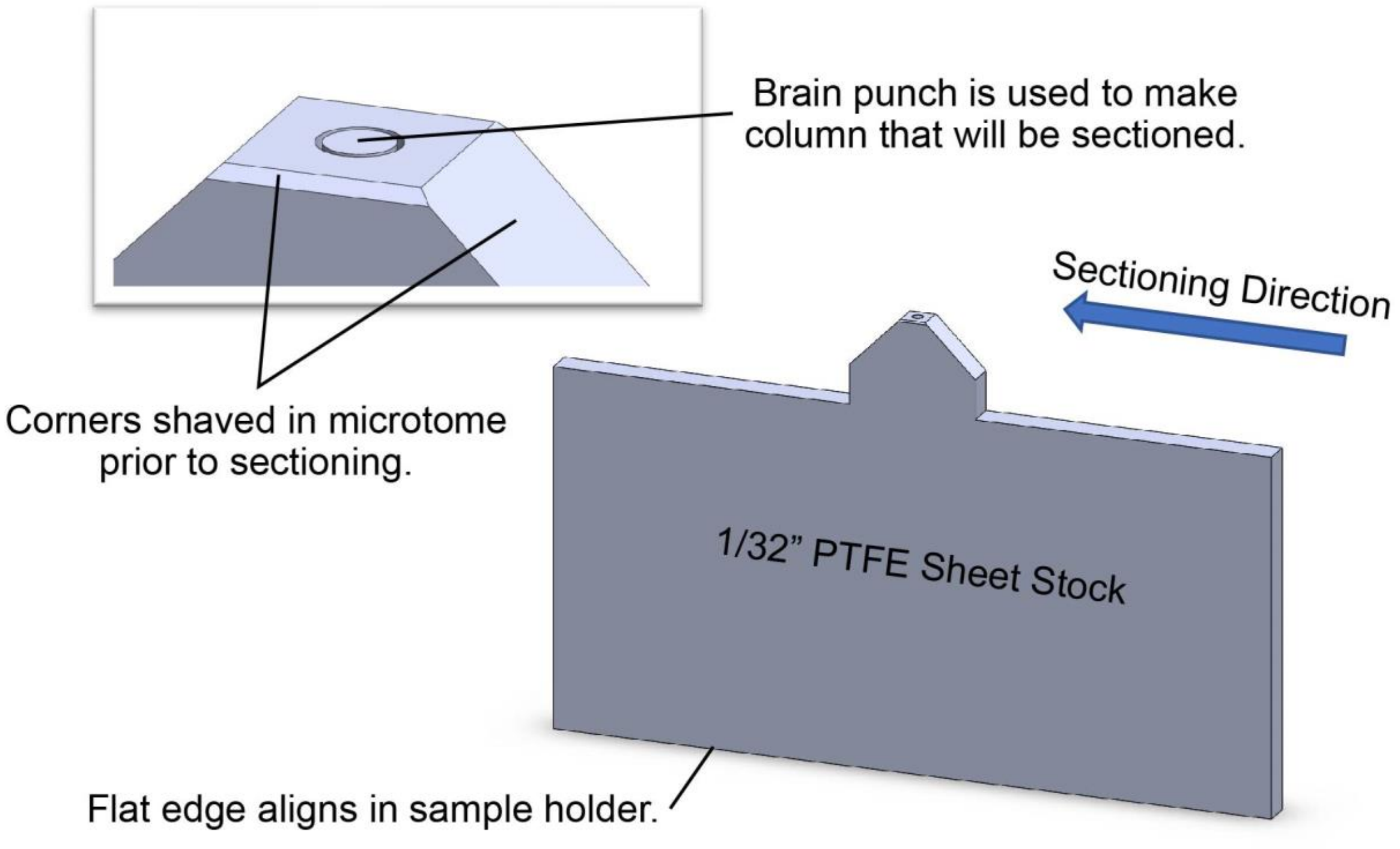

**Fig. S12.**
Procedure and geometry of FSC sample fabrication using the microtome.

| Enrichment | Expected Enthalpy [J/g] | *t* statistic | DoF (Satterthwaite) | p-value (Satterthwaite) | p-value (conservative) |
|---|---|---|---|---|---|
| **Compared to Unexposed — extrapolated enthalpy** = 18.9 ± 0.6 **[J/g]** | | | | | |
| 0.72% (natural) | 19.0 ± 0.7 | 0.48 | 20.8 | 0.635 | 0.640 |
| 5% (typical LEU) | 20.5 ± 0.7 | 6.96 | 20.7 | $8 \times 10^{-7}$ | $2 \times 10^{-5}$ |
| 20% (HEU limit) | 22.9 ± 0.7 | 17.76 | 20.8 | $5 \times 10^{-14}$ | $2 \times 10^{-9}$ |
| 90% (Weapons-grade) | 27.3 ± 0.7 | 36.75 | 20.7 | $2 \times 10^{-20}$ | $7 \times 10^{-13}$ |
| **Compared to 90% (Weapons-grade) — extrapolated enthalpy** = 27.3 ± 0.7 **[J/g]** | | | | | |
| 0.72% (natural) | 19.0 ± 0.7 | −30.95 | 22.0 | $1 \times 10^{-19}$ | $5 \times 10^{-12}$ |
| 5% (typical LEU) | 20.5 ± 0.7 | −25.34 | 22.0 | $9 \times 10^{-18}$ | $4 \times 10^{-11}$ |
| 20% (HEU limit) | 22.9 ± 0.7 | −16.21 | 22.0 | $1 \times 10^{-13}$ | $5 \times 10^{-9}$ |
| **Compared to 20% (HEU limit) — extrapolated enthalpy =** 22.9 ± 0.7 **[J/g]** | | | | | |
| 0.72% (natural) | 19.0 ± 0.7 | −14.74 | 22.0 | $7 \times 10^{-13}$ | $1 \times 10^{-8}$ |
| 5% (typical LEU) | 20.5 ± 0.7 | −9.17 | 22.0 | $6 \times 10^{-9}$ | $2 \times 10^{-6}$ |
| **Compared to 5% (typical LEU) — extrapolated enthalpy =** 20.5 ± 0.7 **[J/g]** | | | | | |
| 0.72% (natural) | 19.0 ± 0.7 | −5.53 | 22.0 | $1 \times 10^{-5}$ | $2 \times 10^{-4}$ |

**Table S1.**

Probabilities that the null hypothesis, assuming the means for samples exposed to different enrichment-level fluences are equivalent, is true. For each comparison, the expected mean enthalpies are listed ± the standard deviation. The calculated t-statistics assume a sample number of 20 for the unirradiated, unexposed samples, and 12 for all irradiated samples. The p-values are calculated using two different methods of estimating the degrees of freedom (DOF) for the two-sample t-test. For the Satterthwaite approximated DOF, the calculated values are shown. For the more conservative estimate, each DOF was 11. Very low p-values by both methods suggest that the null hypothesis is not true, and the alternative, that the means are in fact different, is correct. Of all the tests, only the comparison between natural uranium and unexposed controls is not significantly different.